\documentclass[12pt]{article}
\pdfoutput=1
\usepackage{bbm}
\usepackage{color}
\usepackage{authblk}
\usepackage{latexsym}
\usepackage{epsfig,amssymb,euscript, mathrsfs,cite}
\usepackage{amsmath}
\usepackage{mathrsfs} 
\usepackage{enumerate}
\definecolor{MyDarkBlue}{rgb}{0.15,0.15,0.45}
\usepackage[linktocpage=true]{hyperref}
\hypersetup{
colorlinks=true,
citecolor=MyDarkBlue,
linkcolor=MyDarkBlue,
urlcolor=MyDarkBlue,
pdfauthor={},
pdftitle={},
pdfsubject={hep-th}
}

\newsavebox{\ns}
\newsavebox{\dbrane}
\newsavebox{\dbshort}

\def\be{\begin{equation}}
\def\ee{\end{equation}}
\def\bea{\begin{eqnarray}}
\def\eea{\end{eqnarray}}

\newcommand{\nn}{\notag \\}

\def\eq#1 { \begin{equation} #1 \end{equation} }

\newcommand{\tr}{\mathrm{tr}}

\newlength{\sswidth}

\numberwithin{equation}{section}       

\begin{document}

\begin{titlepage}

\vfill

\begin{flushright}
Imperial/TP/2020/JG/02\\
ICCUB-20-XXX
\end{flushright}

\vfill

\begin{center}
   \baselineskip=16pt
   {\Large\bf Superconformal RG interfaces in holography}
  \vskip 1.5cm
Igal Arav$^1$, K. C. Matthew Cheung$^1$, Jerome P. Gauntlett$^1$\\
Matthew M. Roberts$^1$ and Christopher Rosen$^2$\\
     \vskip .6cm
             \begin{small}\vskip .6cm
      \textit{$^1$Blackett Laboratory, 
  Imperial College\\ London, SW7 2AZ, U.K.}
        \end{small}\\
     \vskip .6cm
             \begin{small}\vskip .6cm
      \textit{$^2$Departament de F\'isica Qu\'antica i Astrof\'isica and Institut de Ci\`encies del Cosmos (ICC), \\
      Universitat de Barcelona, Mart\'i Franqu\`es 1, ES-08028, \\Barcelona, Spain.}
        \end{small}\\
                       \end{center}
\vfill

\begin{center}
\textbf{Abstract}
\end{center}
\begin{quote}
We construct gravitational solutions that holographically describe two different $d=4$
SCFTs joined together at a co-dimension one, planar RG interface and preserving
$d=3$ superconformal symmetry. The RG interface joins $\mathcal{N}=4$ SYM theory on one side with
the $\mathcal{N}=1$ Leigh-Strassler SCFT on the other. We construct a family of such solutions, which
in general are associated with spatially dependent mass deformations on the $\mathcal{N}=4$ SYM side,
but there is a particular solution for which these deformations vanish. We also construct a Janus solution with 
the Leigh-Strassler SCFT on either side of the interface. 
Gravitational solutions associated with superconformal interfaces involving ABJM theory and two $d=3$ $\mathcal{N}=1$ SCFTs with $G_2$ symmetry 
are also discussed and shown to have similar properties, but they also exhibit some new features.
\end{quote}

\vfill

\end{titlepage}

\tableofcontents

\newpage

\section{Introduction}\label{sec:intro}

Conformal defects/interfaces/boundaries are interesting objects to study in quantum field theory
and continue to be a very active topic of investigation (e.g. \cite{Andrei:2018die}). They provide important insights into the
formal structure of quantum field theory, they play a prominent role in string theory and they have a wide variety of applications within the broad context of condensed matter. In this paper we will consider 
renormalisation group (RG) interfaces. An RG interface separates two distinct conformal field theories, CFT$_{UV}$ and CFT$_{IR}$, with CFT$_{IR}$ being the conformal field theory that arises after perturbing CFT$_{UV}$ by a relevant operator and then flowing to the infrared. The RG interface provides
an interesting map between observables in the two theories, as discussed in \cite{Brunner:2007ur,Gaiotto:2012np}, and provides
a novel perspective on the challenging topic of classifying RG flows between CFTs.

Within the context of holography an interesting construction of planar RG interfaces, separating two $d=3$ SCFTs, was studied\footnote{Holographic RG interfaces associated with double trace deformations have also been discussed in \cite{Melby-Thompson:2017aip}.} in \cite{Bobev:2013yra}. 
In particular, strong numerical evidence was provided for the existence of $D=11$ supergravity solutions that describe an RG interface between the $\mathcal{N}=8$ supersymmetric ABJM theory in $d=3$ with $SO(8)$ global symmetry and $\mathcal{N}=1$  SCFTs with $G_2$ global symmetry, which can be obtained from the $SO(8)$ theory via an RG flow. Furthermore, the interface preserves 
an $\mathcal{N}=(0,1)$ superconformal symmetry in $d=2$. We will have more to say about these solutions in this paper, as we summarise below.

The main focus of this paper, however, will be the construction of gravitational solutions that holographically describe planar RG interfaces that separate two $d=4$ SCFTs; specifically, $\mathcal{N}=4$ SYM on one side of the interface with the ``Leigh-Strassler" $\mathcal{N}=1$ SCFT \cite{Leigh:1995ep} on the other. Recall that the Leigh-Strassler (LS) SCFT arises as the IR limit of an RG flow after deforming $\mathcal{N}=4$ SYM by a specific $\mathcal{N}=1^*$ mass deformation which preserves an $SU(2)\times U(1)_R$ global symmetry. More precisely, viewing
$\mathcal{N}=4$ SYM as $\mathcal{N}=1$ SYM coupled to three chiral multiplets, the $\mathcal{N}=1^*$ mass deformation arises by giving
a mass to one of the chiral multiplets. The RG flow, preserving $d=4$ $\mathcal{N}=1$ Poincar\'e supersymmetry as well as the
$SU(2)\times U(1)_R$ global symmetry, were holographically constructed in \cite{Freedman:1999gp}, using a 
truncation of $SO(6)$ $D=5$ gauged supergravity. As such they give rise to exact solutions of type IIB supergravity.

The new type IIB supergravity solutions in this paper, describing RG interfaces separating $\mathcal{N}=4$ SYM with the LS SCFT, 
will also be constructed using a truncation of $SO(6)$ $D=5$ gauged supergravity (slightly enlarged from that used in \cite{Freedman:1999gp}). 
Generically, the RG interface solutions are supported by fermion and boson mass deformations on the $\mathcal{N}=4$ SYM side of the interface, which have non-trivial dependence on the spatial coordinate transverse to the planar interface. 
These deformations preserve $\mathcal{N}=1$ $d=3$ superconformal symmetry as well as an $SU(2)$ global symmetry (i.e. 
they break the
$U(1)_R$ symmetry of the Poincar\'e invariant RG flow). By contrast, on the LS side of the interface
 there are no deformations for any relevant operators. On both sides of the interface there are various operators with spatially dependent expectation values. While this is the generic situation, there is a particularly interesting solution for which the spatially dependent mass deformations on the $\mathcal{N}=4$ SYM side of the interface also vanish.

We construct the new RG interface solutions rather directly as follows. We start with a $D=5$ gravitational ansatz, with $AdS_4$ slices, that manifestly preserves $d=3$ superconformal invariance and then impose boundary conditions on the BPS equations such that on one side of the interface we approach the LS fixed point. By integrating the BPS equations we find solutions that are associated with $\mathcal{N}=4$ SYM on the other side of the interface.
In \cite{Arav:2020obl} we show that these gravitational solutions also arise, slightly indirectly, as limiting solutions of a more general class of Janus\footnote{In this paper a Janus solution will refer to a co-dimension one planar conformal interface that has the same CFT on either side of the interface or the same up to a discrete parity symmetry.
This includes cases where the coupling constants associated with exactly marginal operators vary as one crosses the interface, as in 
e.g. \cite{Bak:2003jk,Clark:2004sb,Clark:2005te,DHoker:2006vfr,DHoker:2006qeo,DHoker:2007zhm,DHoker:2007hhe,Gaiotto:2008sd,Bobev:2020fon}, as well more general situations where spatially dependent relevant operators are active either on the interface
as in \cite{DHoker:2009lky} or with non-vanishing spatial dependence away from the interface as in \cite{Bobev:2013yra,Arav:2018njv,Herzog:2019bom}.}
solutions that are dual to superconformal interfaces with $\mathcal{N}=4$ SYM on both sides of the interface. In the limit in which
the magnitude of the spatially dependent mass deformations on one of the $\mathcal{N}=4$ SYM sides of these Janus solutions 
diverges, we arrive at an RG interface solution with  $\mathcal{N}=4$ SYM on one side and LS on the other side, of the type discussed here.

We will also present an additional type IIB solution which arises as a limiting case of the RG interface solutions. Specifically, when the magnitude of the mass deformation on the $\mathcal{N}=4$ SYM side of the RG interface goes to infinity we obtain a new superconformal Janus interface with the 
LS SCFT now on both sides of the interface, but related by a discrete $R$-parity. More precisely, the $D=5$ gravitational theory has a $\mathbb{Z}_2$ symmetry, which for the 
$\mathcal{N}=4$ SYM $AdS_5$ solution is dual to an $R$-symmetry of 
$\mathcal{N}=4$ SYM. Furthermore, the $D=5$ theory has two additional $AdS_5$
solutions, which we label LS$^\pm$, each dual to the LS SCFT, which are related by the bulk
$\mathbb{Z}_2$ symmetry. Similarly, the Poincar\'e invariant RG flow solutions from the $\mathcal{N}=4$ SYM $AdS_5$ solution
to the LS$^\pm$ $AdS_5$ solutions
are also related by this symmetry. The new Janus solution, which we denote by LS$^+$/LS$^-$, 
has a conformal boundary which approaches the LS$^+$ $AdS_5$ solution on one side of the interface and the LS$^-$ $AdS_5$ solution on the other. Interestingly, the LS$^+$/LS$^-$ Janus solutions are not supported by sources for operators on either side of the interface, but just have operators taking spatially dependent expectation values\footnote{An interesting open question, which is beyond the scope of this paper, is to elucidate 
whether or not there are distributional sources that are located on the interface itself.
For the LS$^+$/LS$^-$ Janus interface, it seems difficult to envisage distributional sources for the scalar operators, while
maintaining conformal symmetry, due to their irrational scaling dimensions.}.
By determining how the expectation value of a certain relevant operator
of the LS theory behaves as the mass deformation on the $\mathcal{N}=4$ SYM side diverges
we are able to identify novel critical exponents which we numerically determine.

Our constructions also include a class of $D=5$ solutions that approach the LS$^\pm$ $AdS_5$ solution on one side of the
interface and are singular on the other side. The singularities, with scalar fields reaching the boundary of the scalar manifold, are similar to
the singularities that arise in Poincar\'e invariant RG flows (e.g. \cite{Freedman:1999gp}). Similar solutions, using
spatially dependent sources, were also found in a bottom up context in \cite{Gutperle:2012hy} (see also \cite{Bobev:2013yra}). In \cite{Gutperle:2012hy} it was suggested that these solutions can be interpreted as being dual to boundary CFTs.  An interesting difference between our solutions and those of \cite{Gutperle:2012hy} is that on the LS side the sources vanish. We leave a further investigation of these solutions, including the precise nature of the singularity in $D=10$ and the corresponding dual interpretation, to future work.

We now return to the solutions discussed in \cite{Bobev:2013yra} which describe RG interfaces between 
$\mathcal{N}=8$ $SO(8)$ ABJM theory in $d=3$ with $\mathcal{N}=1$ SCFTs with $G_2$ global symmetry.
These solutions  were found numerically\footnote{These solutions have also been discussed using a perturbative construction 
in \cite{Kim:2020unz}.}
using a truncation of $D=4$ $SO(8)$ gauged supergravity, as limiting cases of a more general class of Janus solutions
that are holographically dual to superconformal interfaces that separate two copies of the $\mathcal{N}=8$ $SO(8)$ theory on either side of the
interface and preserving, in general $\mathcal{N}=(0,1)$ superconformal symmetry in $d=2$. It was also shown that
these more general Janus solutions are associated with boson and fermion mass deformations, on either side of the interface, that preserve the
$d=2$ $\mathcal{N}=1$ superconformal symmetry. 
Here we will construct the RG interface solutions more directly, 
and hence clarify various aspects of the full moduli space of these solutions as well as elucidate some new properties.

The $D=4$ gravity theory has two $G_2$ invariant $AdS_4$ solutions, labelled $G_2^\pm$, which are dual to two $\mathcal{N}=1$ SCFTs
related by the action of a discrete $CP$ transformation, as we will argue. Correspondingly, there are two different families
of RG interface solutions. In general, the RG interface which separates the $\mathcal{N}=8$ $SO(8)$ theory with one of the $G_2^\pm$ SCFTs, is associated with
spatially dependent mass deformations\footnote{Spatially dependent mass deformations of
ABJM theory that preserve supersymmetry have also been discussed in \cite{Kim:2018qle,Kim:2019kns,Gauntlett:2018vhk,Arav:2018njv}.}
 just on the $\mathcal{N}=8$ side of the interface. Furthermore, we also show that there are again
two particularly interesting solutions for which this source actually vanishes. Additionally, we show that in the limit in which the source diverges, 
in one of the two families of solutions, one obtains the $G_2^+/G_2^-$ Janus solution of \cite{Bobev:2013yra} describing an interface solution which has the
two $G_2$ invariant SCFTs on either side of the interface. We also determine critical exponents associated with how this solution
is approached.

The plan of the paper is as follows. In section \ref{sec2} we discuss the $D=5$ gauge supergravity solutions associated with the interfaces of $d=4$ SCFTs,
while in section \ref{sec3} we discuss the $D=4$ gauge supergravity solutions associated with the interfaces of $d=3$ SCFTs. We end the paper with
some discussion in section \ref{sec:disc}. In appendix \ref{holrena} we include some details of the holographic renormalisation that we use to
analyse the $D=4$ gravity solutions. For the $D=5$ case, which is considerably more involved, details are provided in \cite{Arav:2020obl}.

{\bf Note added}: After this work was finished, two papers appeared \cite{Ooguri:2020sua,Simidzija:2020ukv} which discuss
interfaces of CFTs within holography from a different point of view.

\section{Interfaces of $d=4$ SCFTs}\label{sec2}
\subsection{ $\mathcal{N}=1^*$ one-mass deformations of $\mathcal{N}=4$ SYM}
We begin by recalling a few aspects of homogeneous (i.e. spatially independent) $\mathcal{N}=1^*$ ``one-mass deformations"
of $\mathcal{N}=4$ SYM theory\footnote{The possibility of SCFTs arising from such mass deformations were first discussed in
\cite{Parkes:1982tg} and see \cite{Leigh:1995ep} for a later treatment.}. 
We can view the field content of $\mathcal{N}=4$ SYM in terms of $\mathcal{N}=1$ language as 
a vector multiplet, which includes the gauge-field and the gaugino, coupled to three chiral superfields $\Phi_a$. Under the decomposition of
the $R$-symmetry $SU(3)\times U(1)_1\subset SU(4)_R$ the $\Phi_a$ transform
in the {\bf 3} of $SU(3)\subset SU(4)_R$. The $\mathcal{N}=1^*$ one-mass deformations are obtained by adding mass terms associated
with one of the chiral superfields, say $\Phi_3$. Specifically, we add
to the superpotential $\mathcal{W}_{SYM}$ of $\mathcal{N}=4$ SYM
the term 
\begin{align}\label{spdef}
\Delta\mathcal{W}_{SYM}\sim m ~\tr(\Phi_3^2)\,,
\end{align}
with $m$ a complex and, for homogeneous deformations, constant parameter. 
This deformation gives rise to complex masses for the bosons and fermions in the chiral multiplets.
There is no mass deformation for the gaugino, consistent with preserving $\mathcal{N}=1$ supersymmetry. This homogeneous deformation 
(i.e. with $m$ constant), preserves an
$SU(2)\times U(1)_R$ global symmetry with $U(1)_R$ an $R$-symmetry. The $SU(2)$ factor arises from the
decomposition $SU(2)\times U(1)_2\subset SU(3)$, and the $U(1)_R$ is a diagonal subgroup of $U(1)_1\times U(1)_2$.
Under RG flow this deformation leads to the Leigh-Strassler SCFT in the IR, which has the $SU(2)\times U(1)_R$ global symmetry.
The dual gravitational solutions describing the Poincar\'e invariant RG flow between $\mathcal{N}=4$ SYM and the LS fixed point, were constructed in  
\cite{Freedman:1999gp,Pilch:2000fu}, as we will recall below.  

In the sequel we will construct gravitational RG interface solutions that have $\mathcal{N}=4$ SYM and the LS fixed point on either side of the planar interface. As we will see, the solutions have non-vanishing sources for boson and fermion masses on the $\mathcal{N}=4$ SYM side of the interface
that depend on the spatial direction transverse to the interface, which we take to be $y_3$. In field theory language this means that $m\to m(y_3)$ in \eqref{spdef}. From the analysis of \cite{Anderson:2019nlc} we deduce that this can preserve supersymmetry, provided that we 
include specific $F$ terms in the superpotential. 
This leads to the same
fermion masses, of the form $m \tr \chi_3^2+h.c.$, but deforms the scalar mass term via
$|m|^2  \tr |Z_3|^2  \pm (m' \tr Z_3^2 + \mathrm{h.c.})$,
where $Z_3$ and $\chi_3$ the bosonic
and fermionic components of the superfield $\Phi_3$, respectively. The bosonic mass term $m'$  breaks the $SU(2)\times U(1)_R$ global symmetry of the homogeneous mass deformations down to
$SU(2)$. Moreover, the deformation
will preserve $d=3$ $\mathcal{N}=1$ superconformal symmetry of the interface provided that $m(y_3)\propto 
1/y_3$. 
Further details on these field theory results can be found in \cite{Arav:2020obl}.

\subsection{The $D=5$ gravity model}
We will use a $D=5$ theory of gravity, called the $\mathcal{N}=1^*$ one mass model in \cite{Bobev:2016nua},
that arises as a consistent truncation of $\mathcal{N}=8$ $SO(6)$ gauged supergravity and hence as a consistent
Kaluza-Klein truncation of type IIB supergravity
\cite{Schwarz:1983qr,Howe:1983sra} reduced on a five-sphere. 
This means, by definition, that solutions can be uplifted on the five-sphere to obtain exact supergravity solutions of type IIB
\cite{Lee:2014mla,Baguet:2015sma}.
We will follow the conventions used in \cite{Bobev:2016nua} and in particular use a {\it mostly minus} $(+,-,-,-,-)$ signature for the metric.

The bosonic field content consists of the metric coupled to a complex scalar $z$ and a real scalar $\beta$.
The gravity-scalar part of the Lagrangian takes the form
\begin{align}
\mathcal{L} = -\frac{1}{4}R +3(\partial\beta)^2 +\frac{1}{2}\mathcal{K}_{z\bar{z}}\partial_{\mu}z\partial^{\mu}\bar{z} - \mathcal{P}  \,,
 \end{align}
where $\mathcal{K}_{z\bar{z}}=\partial_z\partial_{\bar z}\mathcal{K}$ and the K\"ahler potential is given by
 \begin{align}\label{kpot}
\mathcal{K}=-4\log(1-z\bar{z})\,.
\end{align}
The scalar potential ${\cal P}$ can be derived from a superpotential-like quantity
\begin{align}\label{superpotlike}
\mathcal{W}=\frac{1}{L}e^{4\beta}(1+6z^2+z^4)+\frac{2}{L}e^{-2\beta}(1-z^2)^2\,,
\end{align}
via 
\begin{align}
   \mathcal{P} = \frac{1}{8}e^{\mathcal{K}}\Big(\frac{1}{6}\partial_{\beta}\mathcal{W}\partial_{\beta}\overline{\mathcal{W}}
+\mathcal{K}^{\bar{z} z}    
      \nabla_{z}\mathcal{W}\nabla_{\bar{z}}\overline{\mathcal{W}} -\frac{8}{3}\mathcal{W}\overline{\mathcal{W}}\Big)\,,
 \end{align}
with $\mathcal{K}^{\bar{z}z}=1/\mathcal{K}_{z\bar{z}}$ and 
$ \nabla_{z}\mathcal{W}\equiv \partial_z\mathcal{W}+\partial_z \mathcal{K}\mathcal{W}$.
As in \cite{Bobev:2016nua} we can write the complex scalar field in terms of two real scalar fields, $\alpha$ and $\phi$, via
\begin{align}
     z &= \tanh \Big[ \frac{1}{2} \big( \alpha
       -i \phi \big) \Big] \,.
    \end{align}
We note that the bosonic part of this  theory is invariant under the $\mathbb{Z}_2$ symmetry, 
\begin{align}\label{zed2}
z\to -z\,.
\end{align}

This model admits an $AdS_5$ vacuum solution, with $z=\beta=0$ and radius $L$, that uplifts to the $AdS_5\times S^5$ solution,
dual to $\mathcal{N}=4$ SYM theory. By analysing the linearised fluctuations of the scalar fields around this solution we deduce
that $\phi$ is dual to a fermion mass operator $\mathcal{O}^{\Delta=3}_\phi$, with conformal dimension $\Delta=3$, while $\alpha$ and $\beta$ are dual to bosonic mass
operators $\mathcal{O}^{\Delta=2}_\alpha$ and $\mathcal{O}^{\Delta=2}_\beta$, both with $\Delta=2$. Schematically, we have\footnote{Recall that the supergravity modes do not capture the Konishi operator $\tr(|Z_1|^2+|Z_2|^2 + |Z_3|^2 )$. }
\begin{align}\label{opfieldmapz}
  \phi &\quad \leftrightarrow \quad \mathcal{O}^{\Delta=3}_\phi=\tr(\chi_3\chi_3+\text{cubic in $Z_a$})+h.c.\,,  \nn
\alpha &\quad \leftrightarrow \quad \mathcal{O}^{\Delta=2}_\alpha=\tr(Z_3^2)+h.c.\,,\nn
\beta &\quad \leftrightarrow \quad \mathcal{O}^{\Delta=2}_\beta=\tr(|Z_1|^2+|Z_2|^2 -2 |Z_3|^2 )\,,
   \end{align}
where $Z_a$ and $\chi_a$ are the bosonic and fermionic components of the chiral superfields $\Phi_a$ that we discussed in the previous subsection.
Notice that this truncation is suitable for studying real mass deformations of $\mathcal{N}=4$ SYM theory.
We also note that for the $\mathcal{N}=4$ SYM $AdS_5$ solution the bulk $\mathbb{Z}_2$ symmetry
\eqref{zed2} can be identified\footnote{More precisely, there is a $\mathbb{Z}_4$ symmetry of $\mathcal{N}=4$ SYM which gives a $\mathbb{Z}_2$ transformation on the bosonic fields.} as being dual to a discrete (internal) $R$-parity transformation of $\mathcal{N}=4$ SYM.

\newcommand\LLS{\tilde L}

The $D=5$ model also admits two other $AdS_5$ solutions, which we label by LS$^\pm$, given by
\begin{align}
z &= \pm i (2-\sqrt{3})\quad \Leftrightarrow \quad  \phi=\mp\frac{\pi}{6}, \quad \alpha=0\,,       \nn
\beta &= -\frac{1}{6} \log(2)\,,\qquad
\LLS= \frac{3}{2^{5/3}} L,
\end{align}
with $\LLS$ the radius of the $AdS_5$ space for both LS$^\pm$ solutions. The two solutions
are related by the bulk $\mathbb{Z}_2$ symmetry \eqref{zed2} of the $D=5$ gravitational theory.
When uplifted to type IIB these fixed point solutions preserve $SU(2)\times U(1)_R$ global symmetry and are each holographically dual
to the $\mathcal{N}=1$ SCFT found by Leigh and Strassler in \cite{Leigh:1995ep}.
By examining the linearised fluctuations of the scalar fields about the LS$^\pm$ $AdS_5$ solutions , we find
that $\alpha$ is dual to an irrelevant operator $\mathcal{O}^{\Delta=2+\sqrt{7}}_{\alpha}$ with conformal dimension $\Delta=2+\sqrt{7}$. The linearised modes involving 
$\phi$ and $\beta$ mix, and after diagonalisation we find modes that are dual to one relevant and one irrelevant operator in the LS SCFT, which we label
$\mathcal{O}^{\Delta=1+\sqrt{7}}_{\phi,\beta}$ and $\mathcal{O}^{\Delta=3+\sqrt{7}}_{\phi,\beta}$ with dimensions
$\Delta=1+\sqrt{7}\sim3.6$ and $\Delta=3+\sqrt{7}$, respectively.

Gravitational solutions for the homogeneous RG flows, preserving $d=4$ Poincar\'e invariance and flowing from the $\mathcal{N}=4$ SYM $AdS_5$ solution in the UV to LS$^+$ (or LS$^-$) $AdS_5$ solution in the IR, were constructed in \cite{Freedman:1999gp,Pilch:2000fu}.
These flows, which preserve $SU(2)\times U(1)_R$ global symmetry, are driven by a supersymmetric source for the relevant fermion mass operator $\mathcal{O}^{\Delta=3}_{\phi}$
and the bosonic mass operator $\mathcal{O}^{\Delta=2}_\beta$ in $\mathcal{N}=4$ SYM. Furthermore, 
these solutions and can be constructed using the $D=5$ gravitational theory after setting the real part of the complex field to zero, 
$\text{Re}(z)=0$ i.e. $\alpha=0$. The interface solutions which we construct in this paper, break the $U(1)_R$ symmetry and as a consequence we need to take $\text{Re}(z)\ne 0$ i.e. $\alpha\ne 0$.
We also note that the solutions flowing to the LS$^+$ 
and the LS$^-$ $AdS_5$ solutions are related by the bulk $\mathbb{Z}_2$ symmetry \eqref{zed2}. We mentioned above
that this bulk symmetry can be identified as an $R$-parity transformation acting on $\mathcal{N}=4$ SYM in the UV and
so, since the LS$^\pm$ solutions are both dual to the same SCFT, we conclude that the $R$-parity is inducing
an automorphism on the LS SCFT itself.

\subsection{BPS equations for conformal interfaces}\label{bpseqs}
The $D=5$ ansatz for the conformal interface solutions is given by
\begin{align}\label{metjanus}
ds_5^2=e^{2A} ds^2(AdS_4)-dr^2\,,
\end{align}
where the function $A$ as well as the scalar fields $\beta,z$ are taken to be functions of $r$ only.
Here $ds^2(AdS_4)$ is the metric on $AdS_4$ of radius $\ell$, given, for example, in Poincar\'e type coordinates by
\begin{align}\label{ads4poinc}
ds^2(AdS_4)=\ell^2\left[-\frac{dx^2}{x^2}+\frac{1}{x^2}\left(dt^2-dy_1^2-dy_2^2\right)\right]\,,
\end{align}
with $0<x<\infty$.
The factor of $\ell$ can be absorbed after redefining $A$, but we find it helpful to
keep it. The $AdS_4$ isometries of the ansatz implies that it generically preserves a $d=3$ conformal symmetry.

We recover the metric on $AdS_5$ with radius $L$ if we set
\begin{align}\label{ads5scan}
e^{A}=\frac{L}{\ell}\cosh\frac{r}{L}\,,
\end{align}
and $-\infty<r<\infty$.
To see this more clearly, one can first change coordinates via $\cosh(r/L)=1/\cos\mu$, with $-\pi/2<\mu<\pi/2$. Then making the additional change of coordinates $y_3=x\sin\mu$, $Z=x\cos\mu$, we obtain the metric for $AdS_5$ written in Poincar\'e coordinates
\begin{align}
ds^2=L^2[-\frac{dZ^2}{Z^2}+\frac{1}{Z^2}\left(dt^2-dy_1^2-dy_2^2-dy_3^2\right)]\,,
\end{align}
with the ranges 
$0<Z<\infty$ and $-\infty<y_3<\infty$. The conformal boundary is located at $Z=0$ and $y_3$ parametrises one of the spatial dimensions of
this boundary. Note that the coordinates $x,\mu$ are polar coordinates constructed from $y_3,Z$. Thus, the conformal boundary of
$AdS_5$ in the coordinates \eqref{metjanus}, \eqref{ads5scan}
consists of three components: $r\to\infty$ and $x\ne 0$, associated with the 
half space parametrised by $(t,y_i)$ with $y_3>0$,  $r\to-\infty$ and $x\ne 0$, associated with the half space parametrised by $(t,y_i)$ with $y_3<0$, and these are joined at the plane $(t,y_i)$ with $y_3=0$, associated with $x=0$.

We are interested in constructing interface solutions that preserve supersymmetry.
Using the supersymmetry transformations and conventions given in \cite{Bobev:2016nua}, 
for the $D=5$ ansatz we are considering,
one can derive the following BPS equations:
 \begin{align}\label{Apbps}
\partial_rA + \frac{i}{\ell} e^{-A} 
- \frac{1}{3}e^{-i\xi+\mathcal{K}/2}\overline{\mathcal{W}} &=0\,,\nn
i\partial_r \xi
-\frac{1}{2} \left(\partial_z\mathcal{K} \partial_\mu z 
          -   \partial_{\bar z} \mathcal{K} \partial_\mu \bar{z}  \right)
          -\frac{i}{3}\text{Im}\left(e^{-i\xi+\mathcal{K}/2}\overline{\mathcal{W}}\right)  &=0\,,\nn
\partial_r z+ \frac{1}{2}e^{-i\xi+\mathcal{K}/2}\mathcal{K}^{z\bar z} \nabla_{\bar z} \overline{\mathcal{W}} &= 0 ,\nn
\partial_r \beta + \frac{1}{12}e^{-i\xi+\mathcal{K}/2}\partial_{\beta}\overline{\mathcal{W}}&=0.
 \end{align}
Here $\xi=\xi(r)$ is a phase that appears in the expression for the Killing spinors. More details of this calculation as well as the explicit form of the preserved 
Killing spinors are given in \cite{Arav:2020obl}. It is also interesting to highlight that, naively, these BPS equations appear to be over constrained  
due to the reality of $A$ and $\beta$. This turns out not to be the case, due to specific form of $\mathcal{W}$ in \eqref{superpotlike},
and we also expand upon this point in more detail in \cite{Arav:2020obl}.

Of most significance here is to note that if the above BPS equations are satisfied then the full equations of motion are satisfied and,
furthermore, after uplifting to type IIB, the ~$D=10$ solutions generically preserve an $\mathcal{N}=1$, $d=3$ superconformal supersymmetry.
We note here that the BPS equations are obviously invariant under the $\mathbb{Z}_2$ symmetry of the theory, $z\to-z$, mentioned earlier.
In addition, they are also invariant under the $\mathbb{Z}_2$ action 
\begin{align}\label{zedtwoCPR}
r\to-r, \quad z\to \bar z,\quad \xi\to -\xi+\pi\,.
\end{align}
Combining these two we also have the symmetry
\begin{align}\label{zedtwoCP}
r\to-r, \quad z\to -\bar z,\quad \xi\to -\xi+\pi\,.
\end{align}
It is worth noting that this last symmetry leaves invariant\footnote{The phase $\xi$ appearing in the Killing spinor changes by this transformation. However, for each of the LS$^\pm$ $AdS_5$ solutions there is twice as much supersymmetry as the interface solutions and the transformation takes one Killing spinor to another one.} each of the two LS$^\pm$ $AdS_5$ solutions, and is dual to a discrete $CP$ symmetry\footnote{There is an analogue of this symmetry for
Poincar\'e invariant flow solutions from the $\mathcal{N}=4$ SYM $AdS_5$ solution to each of the LS$^\pm$ solutions
and one can identify the symmetry in $\mathcal{N}=4$ SYM as a $CP$.} of the LS SCFT.

\subsection{The $\mathcal{N}=4$ SYM/LS RG interface and LS$^+$/LS$^-$ Janus}
We first consider solutions of the form \eqref{metjanus} that describe a conformal RG interface between $\mathcal{N}=4$ SYM and
the LS SCFT. The $D=5$ gravity theory has two $AdS_5$ solutions, LS$^\pm$, related by the $\mathbb{Z}_2$ symmetry \eqref{zed2}
and each dual
to the LS SCFT; for definiteness we focus on LS$^+$.
We therefore want to solve the BPS equations and impose boundary conditions on the ansatz 
\eqref{metjanus} so that as $r\to\infty$, say, we approach the $\mathcal{N}=4$ SYM
$AdS_5$ solution while as $r\to-\infty$ we approach the LS$^+$ $AdS_5$ solution. 

\subsubsection{Holographic renormalisation}
Before describing the solutions, we discuss a few subtleties that arise in determining the sources and expectation values of various
operators in the dual field theory when implementing holographic renormalisation. 
The $\mathcal{N}=4$ SYM side is the more intricate, so we discuss that first. We begin by noting
that we can develop the asymptotic expansion to the BPS equations schematically given, as $r\to\infty$, by
\begin{align}\label{neqfourbcs}
A&=\frac{r}{L}+\dots\,,\nn
\phi&=\phi_{(s)}e^{-r/L}+\dots+{\phi}_{(v)}e^{-3r/L}+\cdots\,,\nn
\alpha&=\alpha_{(s)}\frac{r}{L}e^{-2r/L}+{\alpha}_{(v)}e^{-2r/L}+\cdots\,,\nn
\beta&=\beta_{(s)}\frac{r}{L}e^{-2r/L}+{\beta}_{(v)}e^{-2r/L}+\cdots\,,
\end{align}
with a number of relations amongst the various constant coefficients appearing. For example, the terms
$\phi_{(s)}$, $\alpha_{(s)}$ and $\beta_{(s)}$, which denote the source terms for the dual operators, must satisfy\footnote{Note the bulk scalar
fields are dimensionless.}
$\alpha_{(s)}=-\frac{L}{\ell} \phi_{(s)}$ and $\beta_{(s)}=-\frac{2}{3}\phi_{(s)}^2$.

As $r\to \infty$ we approach a component of the conformal boundary located  
on one side of the interface, with metric $AdS_4$ as in \eqref{ads4poinc}. Thus,
this expansion is naturally suited to
obtaining the sources and expectation values for the various operators when $\mathcal{N}=4$ SYM is placed on $AdS_4$. 
In fact the field theory sources on $AdS_4$ are given by $\phi_{(s)}$, $\alpha_{(s)}$, $\beta_{(s)}$ and we note 
that $\ell\phi_{(s)}$, $\ell^2\alpha_{(s)}$, $\ell^2\beta_{(s)}$ are invariant under Weyl rescalings of the $AdS_4$ radius $\ell$.
Since we are
primarily interested in the associated quantities when the theory is placed on flat space we need to carry out a suitable Weyl transformation,
with Weyl factor $x^2/\ell^2$ acting on  \eqref{ads4poinc}.
A subtlety in this approach, is that the source terms give rise to terms in the conformal anomaly quadratic and quartic 
in the sources
as in \cite{Petkou:1999fv,Bianchi:2001kw} and discussed in detail in \cite{Arav:2020obl},
which needs to be carefully tracked in order to carry out the Weyl transformation in detail.
An alternative, equivalent procedure, is to implement an asymptotic change of coordinates generalising that discussed below \eqref{ads5scan},
so that one approaches the conformal
boundary associated with the theory on flat space. An additional subtlety in carrying out the holographic renormalisation
is that one must specify a number of finite counterterms that are dual to a choice of renormalisation scheme. The details of the scheme that we
employ (which is more general, but consistent with the ``Bogomol'nyi trick" of \cite{Freedman:2013ryh,Bobev:2013cja,Bobev:2016nua}) are discussed in \cite{Arav:2020obl}. 

The upshot of a rather long analysis is the following.
A solution with boundary conditions \eqref{neqfourbcs} is associated with the following sources for $\mathcal{N}=4$ SYM on flat space: 
\begin{align}\label{n4sces}
\frac{\ell\phi_{(s)}}{y_3},\qquad \frac{\ell^2 \alpha_{(s)}}{y_3^2}, \qquad \frac{\ell^2\beta_{(s)}}{y_3^2}\,,
\end{align}
with $y_3>0$,
and the BPS equations imply that
\begin{align}\label{bpsrels}
\alpha_{(s)}=- \frac{L}{\ell} \phi_{(s)}\,,\qquad
\beta_{(s)}=\frac{2}{3}\phi_{(s)}^2\,.
\end{align}
Note in particular, that all sources can be expressed in terms of $\phi_{(s)}$, which we will use in the plots below.

Furthermore, for the associated expectation values of the operators in flat spacetime, we have
\begin{align}\label{nis4vevs1}
\langle\mathcal{O}_{\alpha}\rangle
&=\frac{1}{4\pi GL}\frac{\ell^2}{y_3^2}\Big({\alpha}_{(v)}+\alpha_{(s)}
\log(\frac{y_3}{\ell e^{2\delta_\alpha}})\Big)\,,
\end{align}
which then, along with $\phi_{(s)}$ determines the remaining expectation values via 
\begin{align}\label{nis4vevs2}
\langle\mathcal{O}_{\phi}\rangle&=
\frac{4}{3}\frac{\ell}{y_3}\langle\mathcal{O}_{\beta}\rangle\,\phi_{(s)}-{2L}\frac{1}{y_3} \langle\mathcal{O}_{\alpha}\rangle
-\frac{L}{4\pi G}\frac{\ell}{y_3^3}\phi_{(s)}\,,\nn
\langle\mathcal{O}_{\beta}\rangle
&=\frac{4\ell}{L}\langle\mathcal{O}_{\alpha}\rangle\,\phi_{(s)}-
\frac{(1+4\delta_{\alpha}-4\delta_{\beta})}{2\pi G L}\frac{\ell^2}{y_3^2}\phi_{(s)}^2\,.
\end{align}
Here $\delta_{\alpha}$, $\delta_{\beta}$ are finite counterterms which we have not fixed. 
While the sources transform covariantly under Weyl transformations of the boundary theory, the expectation values do not,
as the presence of the $\log$ terms in these expressions make manifest. In our numerical results below, we will fix $\ell=1$ (as well as $L=1$)
and discuss the values of $\phi_{(s)}$ and ${\alpha}_{(v)}$, which for a definite choice of finite counterterms then gives all of
the sources and expectation values.

We now consider similar issues for the LS$^+$ side of the interface, which turn out to be considerably simpler.
Firstly, since the scalar operators have irrational scaling dimensions there are no finite counterterms that one can add. Secondly, and for similar reasons, the conformal anomaly does not contain
any source terms for the scalar operators. Thirdly, it turns out to be not possible to add sources for the relevant operator
$\mathcal{O}^{\Delta=1+\sqrt{7}}_{\phi,\beta}$ in the LS theory and be consistent with the BPS
equations. To see this latter point one needs to examine the 
linearised solutions to the BPS equations, expanding about the LS$^+$ $AdS_5$ solution as $r\to-\infty$. 
Since we want to approach the LS$^+$ $AdS_5$ solution we need to also demand that there are no source terms
for the two irrelevant operators $\mathcal{O}^{\Delta=2+\sqrt{7}}_{\alpha}$ and $\mathcal{O}^{\Delta=3+\sqrt{7}}_{\phi,\beta}$.
In addition to a universal mode associated with shifts in the coordinate $r$, we then find that there is a single BPS mode
of the form, as $r\to-\infty$, 
\begin{align}\label{lsmodes}
z&=i(2-\sqrt{3})+i\zeta e^{r(1+\sqrt{7})/\LLS}+\dots\,,\nn
\beta&=-\frac{1}{6}\log 2 + b \zeta e^{r(1+\sqrt{7}) /\LLS}+\dots\,,
\end{align}
parametrised by real $\zeta$ and
\begin{align}
b= \frac{1}{18}\left(3+2\sqrt{3} \right)\left( 1+\sqrt{7}\right)\,.
\end{align}
This mode is associated with the relevant operator $\mathcal{O}^{\Delta=1+\sqrt{7}}_{\phi,\beta}$ in the LS$^+$ theory 
acquiring an expectation value. More precisely, for this side of the interface at $r\to-\infty$, which is $y_3<0$ in the flat space boundary, using
\eqref{lsmodes} we can define
\begin{align}\label{lsvevs1}
\langle\mathcal{O}^{\Delta=1+\sqrt{7}}_{\phi,\beta}\rangle
&\propto\left(\frac{\ell}{-y_3}\right)^{1+\sqrt{7}}\zeta\,.
\end{align}
The two irrelevant operators
$\mathcal{O}^{\Delta=2+\sqrt{7}}_{\alpha}$ and $\mathcal{O}^{\Delta=3+\sqrt{7}}_{\phi,\beta}$
also acquire expectation values proportional to $\zeta$.

\subsubsection{The solutions}

\begin{figure}[h!]
\centering
{\includegraphics[scale=0.42]{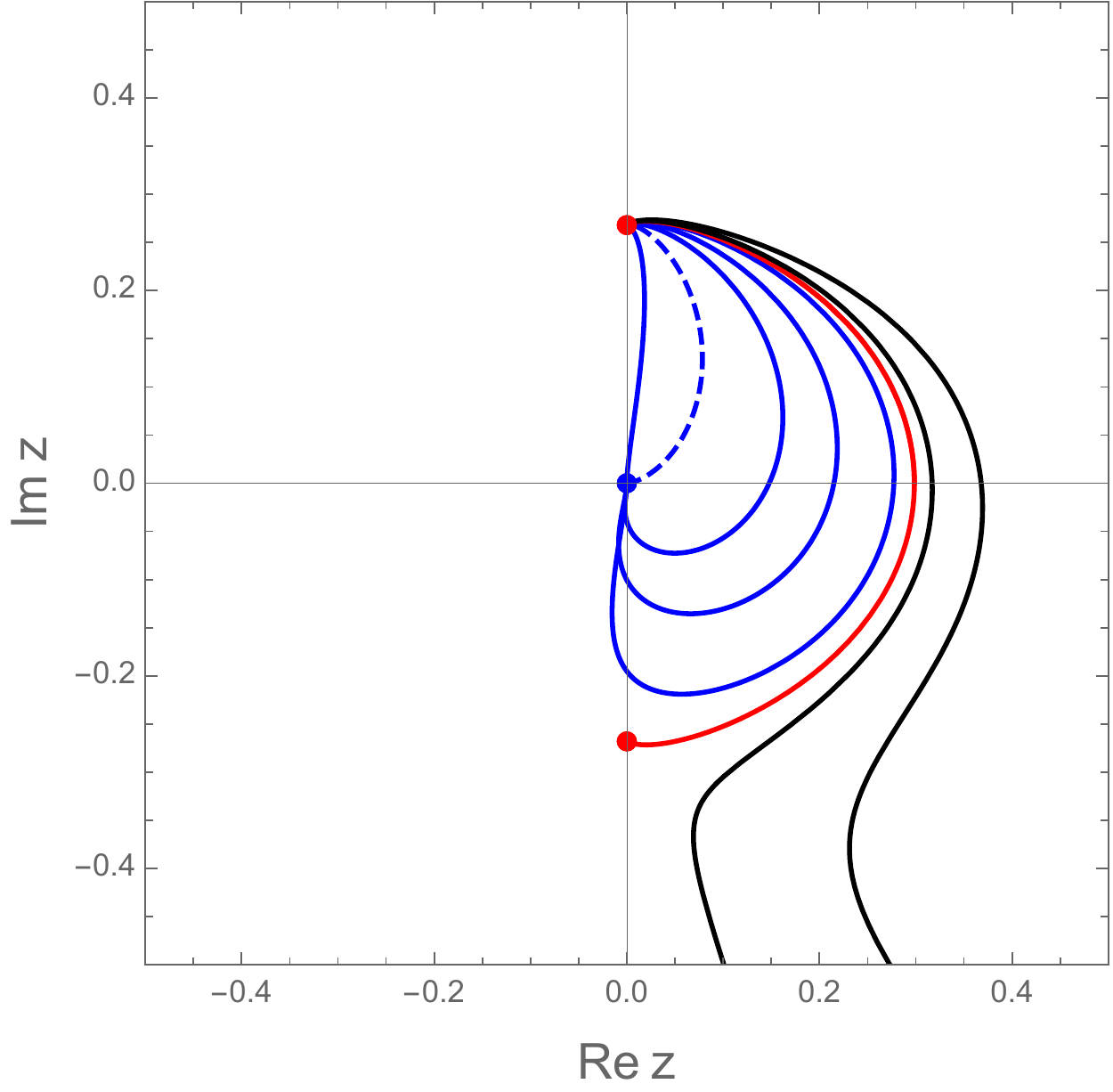}}\\
{\includegraphics[scale=0.27]{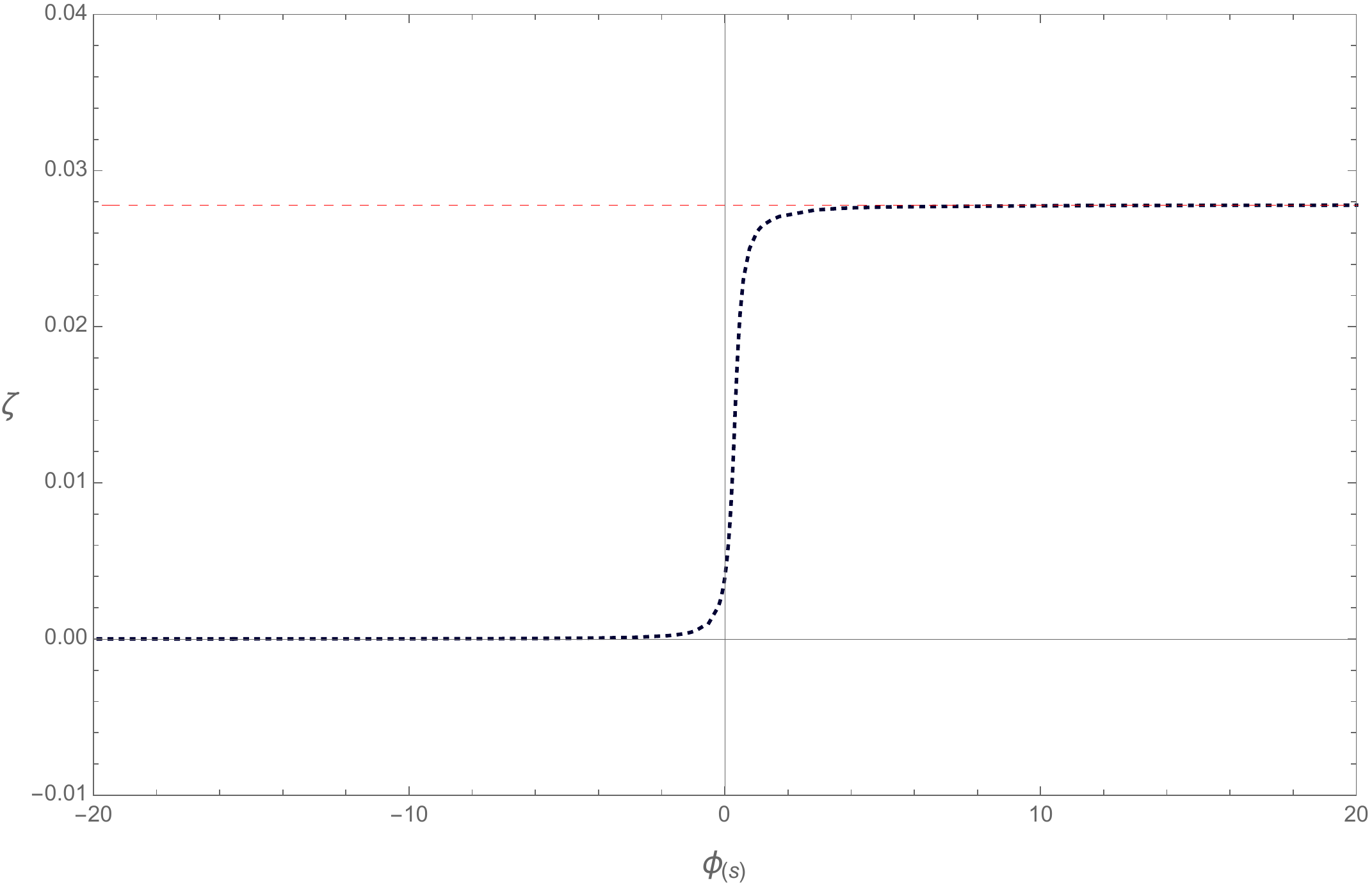}\qquad
\includegraphics[scale=0.33]{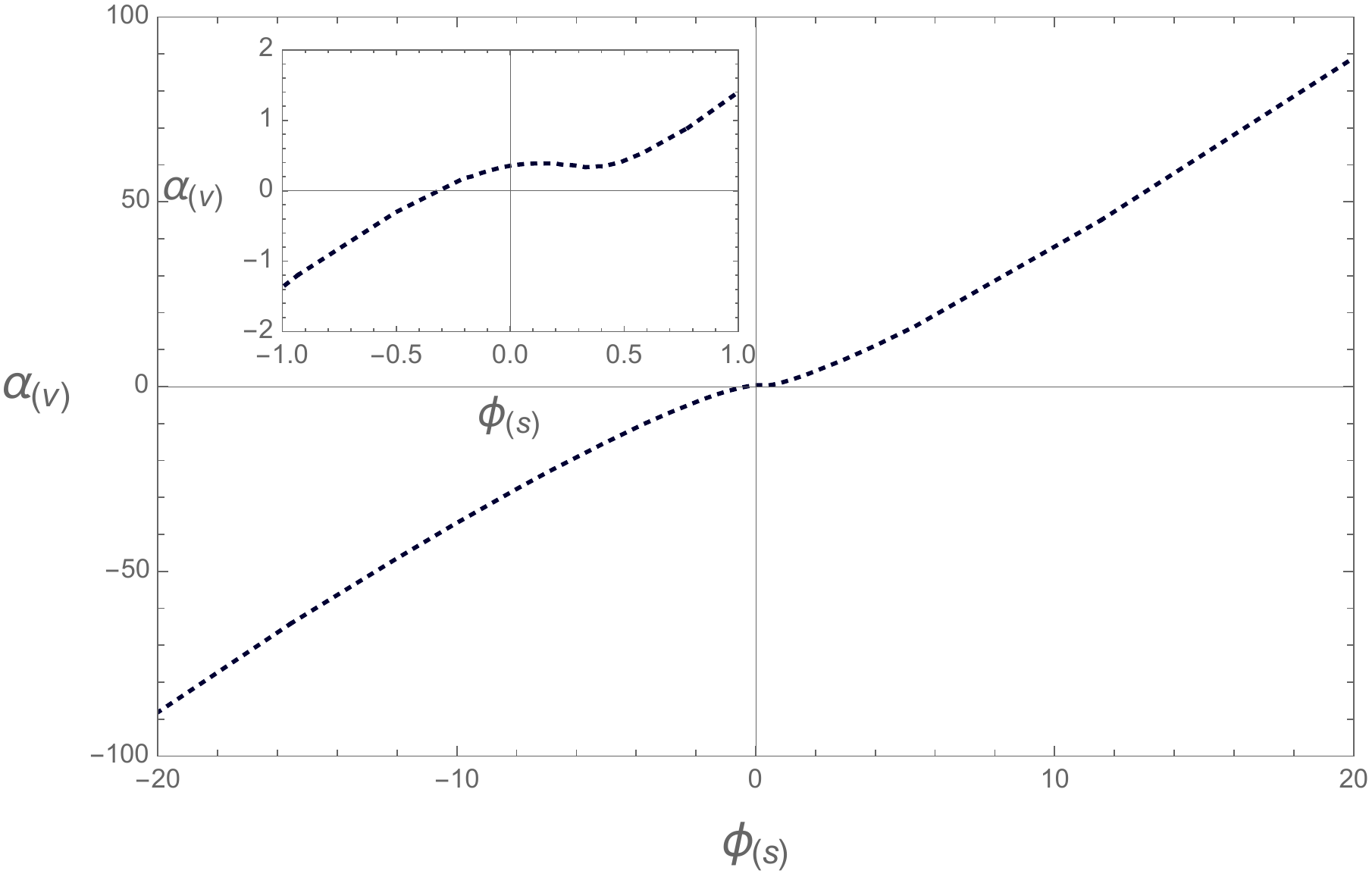}}
\caption{\label{fig:one} The family of $D=5$ BPS solutions is summarised by parametrically plotting the real and imaginary parts of 
the scalar field $z$. The blue dot is the $\mathcal{N}=4$ SYM $AdS_5$ solution and the two red dots
are the two LS $^\pm$ $AdS_5$ solutions. The blue curves are
dual to $\mathcal{N}=4$ SYM/LS$^+$ RG interfaces. 
For these solutions, the bottom panels plot $\zeta$ and $\alpha_{(v)}$, which fix the expectation values on the LS$^+$ and 
$\mathcal{N}=4$ SYM side, respectively, as a function of $\phi_{(s)}$ which fix the
sources on the $\mathcal{N}=4$ SYM side.
The dashed blue line in the top panel is the RG interface solution
for which all sources vanish on the $\mathcal{N}=4$ SYM side of the interface. As $\phi_{(s)}\to+\infty$ one approaches the 
LS$^+$/LS$^-$ Janus solution (red curve). The black curves become singular at $|z|=1$.
 }
\end{figure}

\begin{figure}[h!]
\centering
\raisebox{-0.5\height}{\includegraphics[scale=0.4]{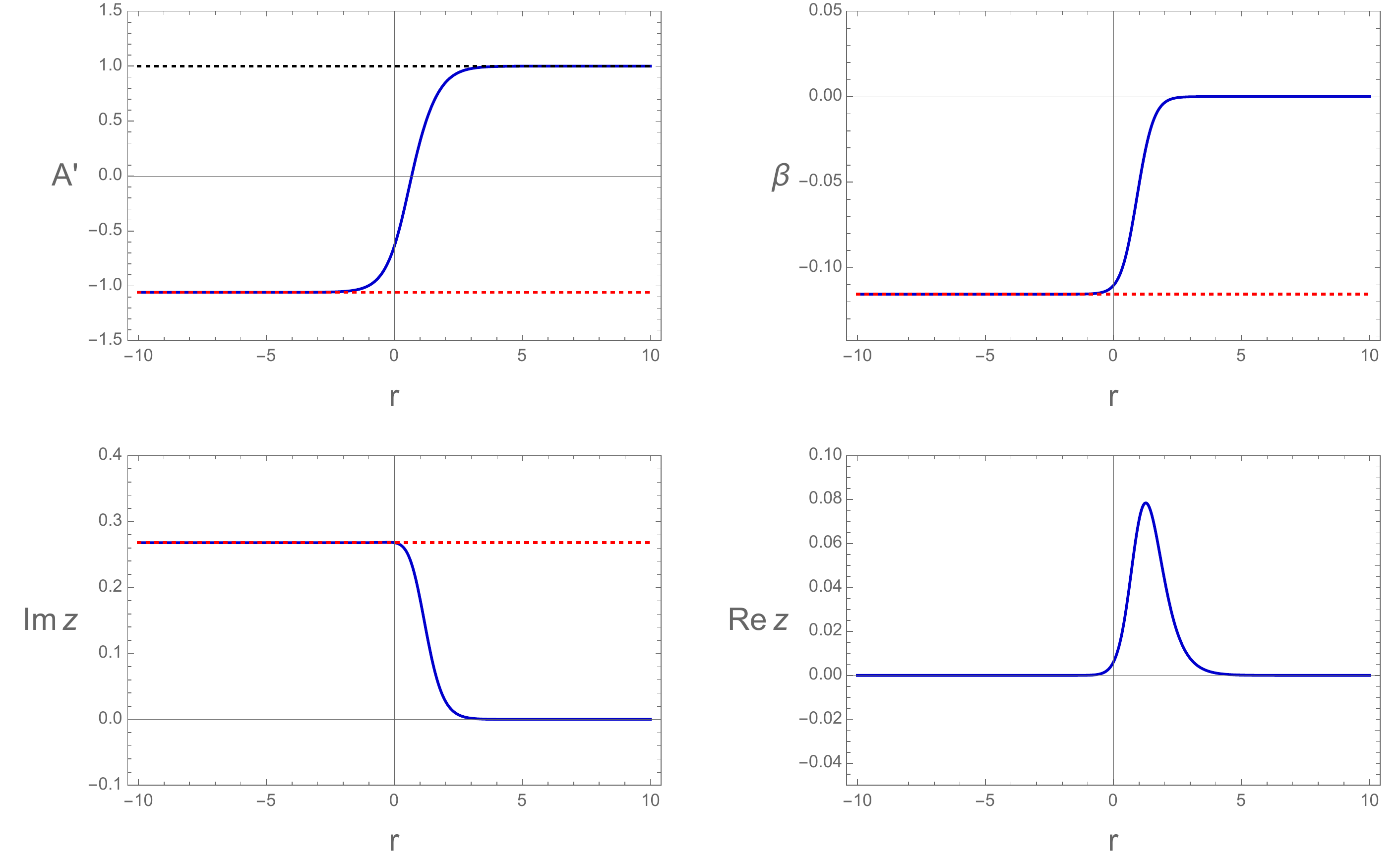}}
\caption{\label{fig:three} The BPS solution for the dashed blue curve in figure \ref{fig:one},
describing an $\mathcal{N}=4$ SYM/LS RG interface, for which 
all sources vanish i.e. $\phi_{(s)}=0$. We have plotted $A'=dA/dr$ as well
as the three scalar functions as a function of $r$. The LS$^+$ $AdS_5$ solution is approached at $r\to-\infty$ while the $\mathcal{N}=4$ SYM $AdS_5$ solution is approached at $r\to\infty$. The red dashed lines, which give the associated values of the LS$^+$ $AdS_5$ solution, have been added to guide the eye.}
\end{figure}

\begin{figure}[h!]
\centering
\raisebox{-0.5\height}{\includegraphics[scale=0.17]{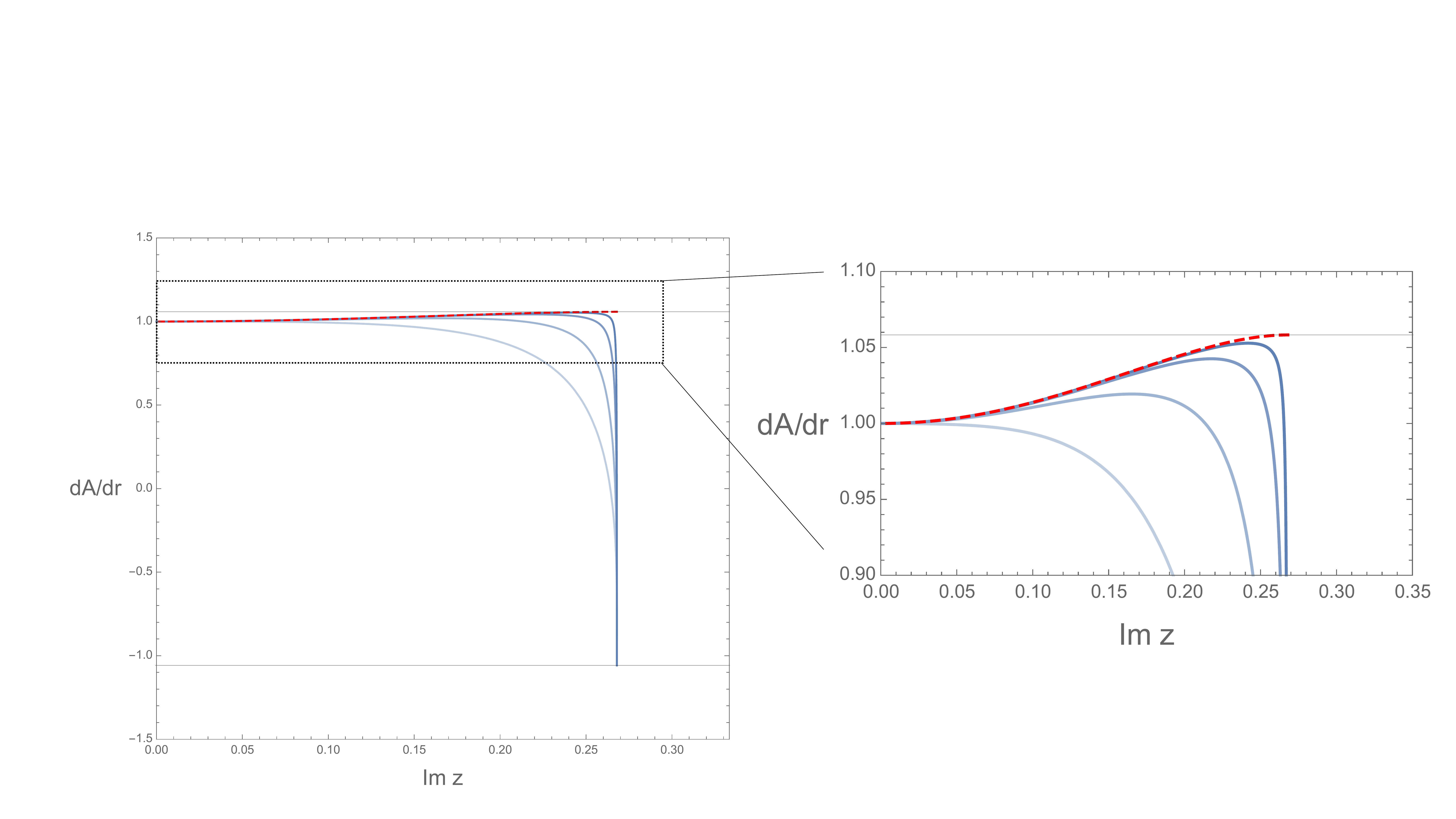}}\\
{\includegraphics[scale=0.4]{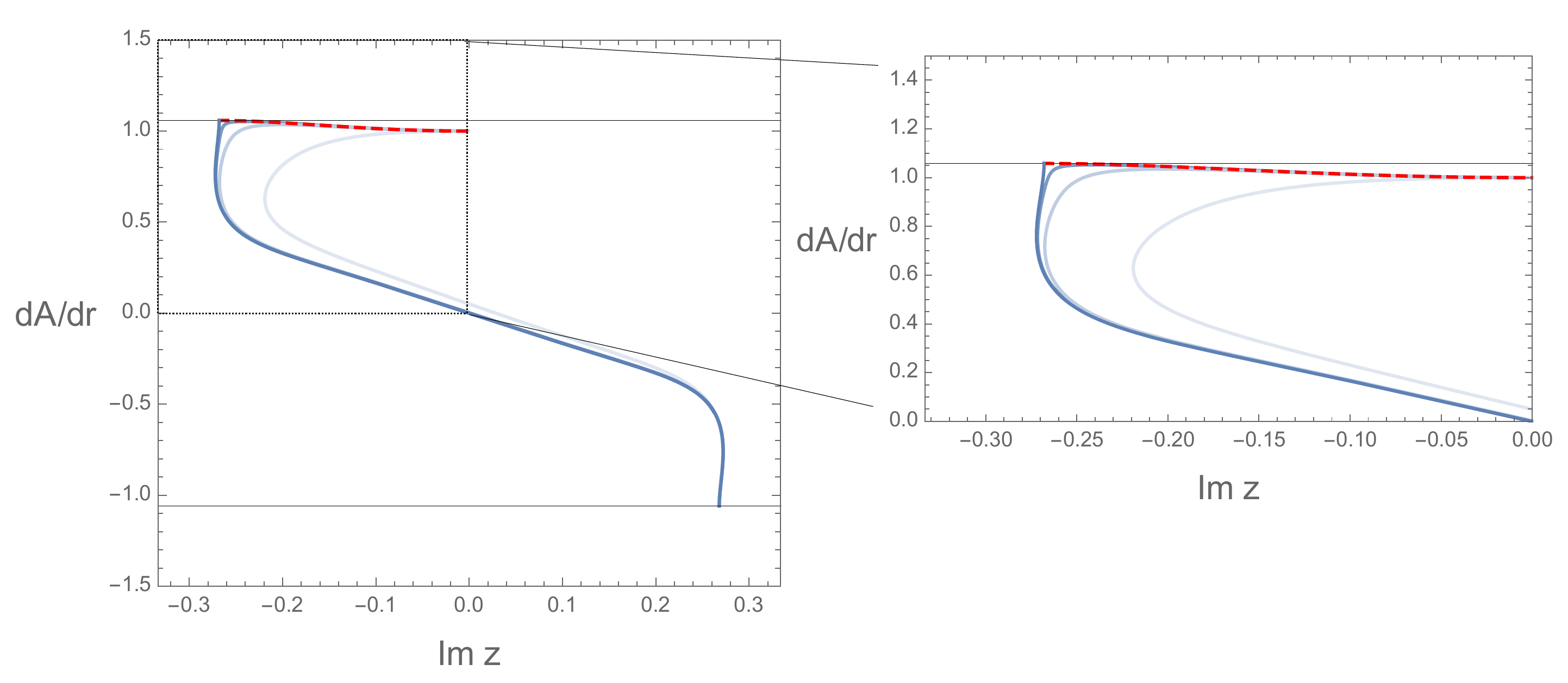}}
\caption{\label{fig:four} We display the limiting behaviour of $\mathcal{N}=4$ SYM/LS RG interface solutions of figure \ref{fig:one}
using parametric plots of $A'$ versus $\text{Im}(z)$. As $\phi_{(s)}\to -\infty$ the 
solutions in figure \ref{fig:one} approach a vertical blue line. In this limit (top panel) the solutions approximate two solutions, the
dashed red line, which is the Poincar\'e invariant RG flow solution from $\mathcal{N}=4$ SYM to the LS$^+$ fixed point, joined
with the vertical blue line, which is the LS$^+$ fixed point itself. As $\phi_{(s)}\to +\infty$ the solutions in figure \ref{fig:one} approach the red curve in
figure \ref{fig:one}. In this limit (bottom panel) 
the solutions approximate two solutions, the
dashed red line, which is the Poincar\'e invariant RG flow solution from $\mathcal{N}=4$ SYM to the LS$^-$ fixed point, joined
with the dark blue line which is the LS$^+$/LS$^-$ Janus solution.}
\end{figure}

We have numerically constructed RG interface solutions by starting with the LS$^+$ side at $r=-\infty$, shooting out with the mode associated with
$\langle\mathcal{O}^{\Delta=1+\sqrt{7}}_{\phi,\beta}\rangle$, parametrised by $\zeta$, and then seeing where one ends up at $r=\infty$.
The main results are presented in figures \ref{fig:one}-\ref{fig:four}. There is another set of physically equivalent solutions that
start with LS$^-$ side at $r=-\infty$, which can be obtained using the $\mathbb{Z}_2$ symmetry \eqref{zed2}, which we won't explicitly discuss.

Figure \ref{fig:one} provides a parametric plot of the real and imaginary parts of the scalar field, $z$, for the solutions we have found.
The blue dot at the origin is the $\mathcal{N}=4$ SYM $AdS_5$ solution while the two red dots are the two LS$^\pm$ $AdS_5$ solutions, related by the $\mathbb{Z}_2$ symmetry \eqref{zed2}. 
The blue curves are a one parameter family of RG interface solutions with $\mathcal{N}=4$ SYM theory on one side ($y_3>0$)
and LS$^+$ on the other ($y_3<0$).
We have also plotted in the bottom left panel $\zeta$, which gives
the expectation values of the LS SCFT, as in e.g. \eqref{lsvevs1},
as a function of  $\phi_{(s)}$, which we recall fixes the fermion mass deformation
as well as all other sources on the $\mathcal{N}=4$ SYM theory side via \eqref{n4sces},\eqref{bpsrels}. 
Similarly in the bottom right panel we have plotted $\alpha_{(v)}$ which, along with $\phi_{(s)}$, fixes the expectation value
on the $\mathcal{N}=4$ SYM theory side, as a function of $\phi_{(s)}$.
The RG interface solutions exist in the range $-\infty <\phi_{(s)}< \infty$ with $0<\zeta<\zeta_{crit}\approx 0.0281$. When $\phi_{(s)}\to +\infty$ 
(and $\zeta\to \zeta_{crit}$) the solutions approach the red curve while
when $\phi_{(s)}\to -\infty$ (and $\zeta\to 0$) they approach a vertical line along the imaginary $z$ axis.

We next note that the lower panels in figure \ref{fig:one} clearly reveal the existence of an RG interface solution for which 
 $\phi_{(s)}=0$. This means that all sources on the $\mathcal{N}=4$ SYM side vanish, and since the sources always vanish on
 the LS$^+$ side, remarkably this is an RG interface solution that has vanishing sources away from the interface.
  For this special solution, marked by the 
dashed blue line in figure 
\ref{fig:one}, we can determine the expectation values of the operators in the two SCFTS.
On the LS$^+$ side we find $\zeta\approx 0.0040$. For the $\mathcal{N}=4$ side, recalling
from \eqref{bpsrels}-\eqref{nis4vevs2} that the expectation values of the scalar operators are all determined by  
$\alpha_{(v)}$ and $\phi_{(s)}$, we note that $\alpha_{(v)}=0.3553$.

The general behaviour of the radial functions for all of the $\mathcal{N}=4$ SYM/LS$^+$ RG interface solutions 
(blue curves in figure \ref{fig:one}) have a similar form. As a representative example, in figure \ref{fig:three} we display the metric and scalar functions for the special source-free solution with $\phi_{(s)}=0$.

We next consider how the RG interface solutions behave as $\phi_{(s)}\to -\infty$ (and $\zeta\to 0$), when they approach a vertical blue line in figure \ref{fig:one}. 
In this limit, one can show that the solutions have a region, on the $\mathcal{N}=4$ SYM side, that closely approaches the 
Poincar\'e invariant RG flow solution from $\mathcal{N}=4$ SYM to the LS$^+$ fixed point, as one might anticipate. 
To make this precise\footnote{In order to recover the expectation value of scalar operators of the Poincar\'e invariant
RG flow one needs to carefully track log terms that arise due to the conformal anomaly - see \cite{Arav:2020obl}.} we can
reinstate $\ell$ and then hold $\phi_{(s)}$ fixed while taking $\ell\to\infty$, so that we are solving the BPS equations on the
$\mathcal{N}=4$ SYM side such that the $\frac{1}{\ell}$ term in \eqref{Apbps} is not playing a significant role. With $\ell=1$, as we have 
assumed in our numerics, we can see the approach to the Poincar\'e invariant solution by simply plotting parametrically
the behaviour of $A'$ with respect to the imaginary part of $z$ (recall that in the Poincar\'e invariant RG solution the real part of $z$ vanishes) as we have done in figure \ref{fig:four}.
Interestingly, in the limit that as $\phi_{(s)}\to -\infty$ we can analyse the way in which 
$\zeta\to 0$ on the LS$^+$ side. This gives rise to the following critical exponent, which from our numerics is given by
\begin{align}\label{cexp1}
\zeta\sim  |\phi_{(s)}|^{-\gamma},\qquad\qquad \gamma \approx1.6457\,.
\end{align}
Recalling that $\zeta$ is giving the expectation value of an operator with conformal dimension $1+\sqrt{7}$ as in
\eqref{lsvevs1}, it seems highly likely that the exact critical exponent is $-1+\sqrt{7}$ and it would be interesting to prove this.

We now consider what happens to the RG interface solutions as $\phi_{(s)}\to+\infty$, when 
$\zeta\to \zeta_{crit}\approx 0.0281$, as in the lower left panel of figure \ref{fig:one}. In this limit the blue curves in 
figure \ref{fig:one} approach the red curve which is a new type of interface solution. Indeed, the red curve describes
a Janus solution with LS$^+$ on one side of the interface and LS$^-$ on the other.
Interestingly, on both sides of the Janus interface we have vanishing sources.
We can also show that this solution is invariant under the $\mathbb{Z}_2$ symmetry \eqref{zedtwoCPR}.
The way in which the RG interface solutions approach this LS$^+$/LS$^-$ Janus is also interesting.
From figure \ref{fig:one} one might expect that on the $\mathcal{N}=4$ SYM side of the interface ($r\to\infty$), the solution
starts to approach the Poincar\'e invariant RG flow solution from $\mathcal{N}=4$ SYM to the LS$^-$ fixed point. Indeed, this
is the case, with the limiting solutions behaving analogously to those in \ref{fig:four}. Focussing now on the LS$^+$ side we obtain
another critical exponent:
\begin{align}
\zeta_{crit}-\zeta\sim  \phi_{(s)}^{-\gamma},\qquad\qquad \gamma \approx1.6459\,,
\end{align}
and again we believe this to be exactly $-1+\sqrt{7}$.

Figure \ref{fig:one} also shows that there is a one parameter family of solutions which approach LS$^+$ as $r\to-\infty$, and then
approach a singular behaviour, with $|z|\to 1$, at some finite value of $r$. These solutions can be 
characterised by the expectation value of the operator $\langle\mathcal{O}^{\Delta=1+\sqrt{7}}_{\phi,\beta}\rangle$ in the LS SCFT and have $\zeta>\zeta_{crit}$, seemingly existing for arbitrary large values of $\zeta$. Although not plotted in figure \ref{fig:one} there are also singular solutions starting at  LS$^+$ with $\zeta<0$ and hitting
a singularity at $|z|=1$. These solutions describe configurations of the LS SCFT when placed on 
a half space without sources and with non-vanishing expectation values. Similar solutions were discussed in 
\cite{Gutperle:2012hy} in a bottom up context where they were interpreted as being dual to boundary CFTs. 
An important difference, however, is that the solutions in \cite{Gutperle:2012hy} were supported by non-vanishing sources.  We also note that
the singularity of the solutions we have constructed are similar
to the singularities that arise in Poincar\'e invariant RG flows (e.g. \cite{Freedman:1999gp}) and it would be interesting to investigate this further.

\section{Interfaces of $d=3$ SCFTs}\label{sec3}
We now consider interfaces between superconformal field theories in $d=3$. Strong numerical evidence for these solutions was 
provided in \cite{Bobev:2013yra} where they arose as limiting solutions of a larger family of Janus solutions. 
Here we will construct the solutions more directly and hence determine the moduli space of these solutions. We will also
elucidate some of the physical properties of these solutions, analogous to the $D=5$ case.

\subsection{The $D=4$ gravity model}
We will use the conventions of \cite{Bobev:2013yra} and, in particular, we now switch to a {\it mostly plus signature} $(-,+,+,+)$. 
The $D=4$ 
theory couples a metric to a complex scalar $z$ with Lagrangian given by\footnote{We have set $g=(\sqrt{2}L)^{-1}$ in \cite{Bobev:2013yra}.}
\begin{align}\label{bulklag}
\mathcal{L}=\frac{1}{2}R-\mathcal{K}_{z\bar{z}}\partial_{\mu}z\partial^{\mu}\bar{z}-\mathcal{P}\,,
\end{align}
where 
$\mathcal{K}_{z\bar{z}}=\partial_z\partial_{\bar z}\mathcal{K}$
and the K\"{a}hler potential is given by
\begin{align}
\mathcal{K}=-7\log(1-z\bar{z})\,.
\end{align}
The potential $\mathcal{P}$ is defined to be
\begin{align}
\mathcal{P}=e^{\mathcal{K}}(\mathcal{K}^{z\bar{z}}\nabla_{z}\mathcal{V}\nabla_{\bar{z}}\mathcal{\overline{V}}-3\mathcal{V}\mathcal{\overline{V}})\,,
\end{align}
where $\mathcal{K}^{z\bar{z}}$ is the inverse of $\mathcal{K}_{z\bar{z}}$ and
\begin{align}
\mathcal{V}&=\frac{1}{L}(z^7+7z^4+7z^3+1)\,,\nn
\nabla_{z}\mathcal{V}&=\partial_{z}\mathcal{V}+\mathcal{V}\partial_z \mathcal{K}\,.
\end{align}
It will be convenient to split $z$ into its real and imaginary parts as follow
\begin{align}
z=X+iY\,.
\end{align}

This model admits five $AdS_4$ fixed points. First, there is the vacuum $AdS_4$ solution, with $z=0$, which uplifts to the maximally supersymmetric $AdS_4\times S^7$ solution dual to the $SO(8)$ invariant ABJM theory\footnote{For all of the solutions we discuss we can also take 
the quotient $S^7/\mathbb{Z}_k$ associated with other ABJM theories. This will not break supersymmetry for $k=1,2$.}.
Next, there are two $AdS_4$ fixed points, labelled $G_2^\pm$, which are dual to $\mathcal{N}=1$ SCFTs in $d=3$ with $G_2$ global symmetry 
and related to each other by a discrete $CP$ symmetry, as we explain below. Finally, there 
are also two $AdS_4$ fixed points, labelled $SO(7)^\pm$, which do not preserve any supersymmetry, and will not be relevant for the interface
solutions we discuss in this section.

We first consider the linearised spectrum about the ABJM $AdS_4$ solution, with $z=0$. Fluctuations of the scalar field $z=X+iY$ have mass squared equal to $-2/L^2$ and hence correspond to operators with $\Delta=1,2$ depending on how we quantise. To preserve $\mathcal{N}=8$ supersymmetry we require $Y$ is dual to a fermion mass operator $\mathcal{O}^{\Delta=2}_Y$, with conformal dimension $\Delta=2$, while
$X$ is dual bosonic mass operator $\mathcal{O}^{\Delta=1}_X$, with conformal dimension $\Delta=1$ (e.g. \cite{Bobev:2013yra}). This can be
implemented by adding suitable boundary terms to the action. In appendix \ref{holrena} we have summarised the supersymmetric
holographic renormalisation scheme we employ which incorporates these boundary terms as well as specific finite counterterms, which allows us
to obtain the sources and expectation values of the dual fermion and boson mass operators.

We next consider the two $G_2^\pm$ $AdS_4$ solutions given by 
\begin{align}
z =z_{(G_2^\pm)}\equiv\frac{\sqrt{(2\sqrt{3}-3)}\pm i}{\sqrt{5}+\sqrt{(2\sqrt{3}+3)}}
\,,\qquad
\LLS=\frac{5^{5/4}}{2^{5/4}3^{9/8}}L\,,
\end{align}
with $\LLS$ the radius of the $AdS_4$ space for both $G_2^\pm$ solutions. 
By examining the linearised fluctuations of the scalar field $z$ about these solutions, we find 
modes associated with two irrelevant operators in the $G_2^\pm$ SCFTs, 
$\mathcal{O}^{\Delta=1+\sqrt{6}}_{z}$ and $\mathcal{O}^{\Delta=2+\sqrt{6}}_{z}$
with conformal dimensions
$\Delta= 1+\sqrt{6} \sim3.45$ and $\Delta=2+\sqrt{6}$, respectively.

Gravitational solutions for the homogeneous RG flows, preserving $d=3$ Poincar\'e invariance and flowing from maximally supersymmetric $d=3$ SCFT in the UV to the $G_2^\pm$ fixed points in the IR, were studied in \cite{Ahn:1999zy,Ahn:2000mf,Ahn:2001kw,Bobev:2009ms}.
These flows are driven by a supersymmetric source for the relevant fermion and bosonic mass operators $\mathcal{O}^{\Delta=2}_{Y}$
and $\mathcal{O}^{\Delta=1}_X$, respectively, and preserve $\mathcal{N}=1$ supersymmetry as well as the $G_2$ global symmetry.
It is also interesting to point out that the two $G_2^\pm$ SCFTs are related by a discrete $CP$ symmetry. To see this, we recall that
the ABJM field theory has such a symmetry, which also involves exchanging the two gauge groups of the theory  \cite{Aharony:2008ug}.
The RG flow from ABJM to the $G_2^+$ fixed point is mapped to the RG flow from ABJM to the $G_2^-$ fixed point under this $CP$ action and
hence the two fixed points themselves are similarly mapped into each other.

\subsection{BPS equations for conformal interfaces}\label{bpseqs2}

The $D=4$ ansatz for the conformal interface solutions is given by
\begin{align}\label{metjanusf}
ds_4^2=e^{2A} ds^2(AdS_3)+dr^2\,,
\end{align}
where the function $A$ as well as the scalar field $z$ are taken to be functions of $r$ only.
Here $ds^2(AdS_3)$ is the metric on $AdS_3$ of radius $\ell$,  and hence the ansatz  generically preserves a $d=2$ 
conformal symmetry. The metric on $AdS_4$ with radius $L$ is recovered by setting 
$e^{A}=\frac{L}{\ell}\cosh\frac{r}{L}$ with $-\infty<r<\infty$.

We now impose the conditions for supersymmetry. Using the supersymmetry transformations\footnote{Note that we should identify
$\mathcal{V}^{here}=\frac{1}{2}W^{there}$. We have also changed the projection in (A.12) of \cite{Arav:2018njv} to $\Gamma^{\hat{t}\hat{y}}\hat\epsilon=+\hat\epsilon$ which implies that in the BPS equations of (A.24) of \cite{Arav:2018njv} we should take $x\to -x$. }
 given in \cite{Arav:2018njv}, 
for the $D=4$ ansatz we are considering, we obtain the following BPS equations:
\begin{align}\label{bpsd4}
\partial_r A-\frac{i}{\ell}e^{-A}&=e^{i\xi}e^{\mathcal{K}/2}\mathcal{V}\,,\nn
\partial_r z&=-e^{-i\xi}e^{\mathcal{K}/2}\mathcal{K}^{z\bar{z}}\nabla_{\bar{z}}\mathcal{\overline{V}}\,,\nn
\partial_r\xi+3\mathcal{A}_{r}&=-e^{\mathcal{K}/2}\text{Im}(\mathcal{V}e^{i\xi})\,.
\end{align}
where $\mathcal{A}_r=\frac{i}{6}(\partial_z \mathcal{K}\partial_r z-\partial_{\bar z} \mathcal{K}\partial_r \bar z)$.
Here $\xi=\xi(r)$ is a phase that appears in the expression for the Killing spinors. If the
BPS equations are satisfied then the full equations of motion are satisfied and,
furthermore, after uplifting to $D=11$ supergravity, the solutions generically preserve an $\mathcal{N}=1$ superconformal supersymmetry in $d=3$.
Notice that the BPS equations are invariant under the $\mathbb{Z}_2$ action 
\begin{align}\label{zedtwo4}
r\to-r, \quad z\to \bar z,\quad \xi\to -\xi+\pi\,.
\end{align}
analogous to what we saw in the $D=5$ case in \eqref{zedtwo4}. 
This symmetry takes the $G_2^+$ $AdS_4$ solution to the $G_2^-$ $AdS_4$ solution
and hence can be identified with the $CP$ transformation mentioned above. 
Note that the $D=4$ theory does not have a discrete symmetry acting on the scalar manifold, analogous to the $z\to -z$ symmetry that we saw
for the $D=5$ model.

\subsection{The ABJM/$G_2^\pm$ RG interfaces and $G_2^+$/$G_2^-$ Janus}
We first consider solutions of the form \eqref{metjanusf} that describe a conformal RG interface between ABJM theory and
one of the two $G_2^\pm$ SCFTs. We therefore want to solve the BPS equations and impose boundary conditions on the ansatz 
\eqref{metjanusf} so that as $r\to\infty$, say, we approach the $\mathcal{N}=8$ $AdS_4$ solution while as 
$r\to-\infty$ we approach either the $G_2^+$ or the $G_2^-$ invariant $AdS_4$ solutions.
It is important to emphasise that the lack of a discrete symmetry of the BPS equations that just acts on the scalars (analogous to the $z\to-z$ symmetry we saw in the $D=5$ model), means that these two types of interface are now physically distinct. Indeed we 
will explicitly see this in the behaviour of the solutions.

\subsubsection{Holographic renormalisation}
We now discuss some features of holographic renormalisation, starting with the $\mathcal{N}=8$ ABJM side of the RG interface.
As $r\to\infty$ we can develop the asymptotic expansion to the equations of motion schematically given by
\begin{align}
A&=\frac{r}{L}+\cdots\,,\nn
X&=X_{(1)}e^{-r/L}+{X}_{(2)}e^{-2r/L}+\cdots\,,\nn
Y&=Y_{(1)}e^{-r/L}+{Y}_{(2)}e^{-2r/L}+\cdots\,.
\end{align}
Using the boundary counterterms summarised in appendix \ref{holrena},
we can then identify the sources dual to the operators $\mathcal{O}^{\Delta=1}_X$ and
$\mathcal{O}^{\Delta=2}_Y$ on the $AdS_3$ boundary to be $X_{(s)}$ and $Y_{(s)}$, respectively,
where
\begin{align}
X_{(s)}&=-{X}_{(2)}+3X_{(1)}^2-3Y_{(1)}^2\,,\nn
Y_{(s)}&=Y_{(1)}\,.
\end{align}
We note that $\ell^2X_{(s)}$ and $\ell Y_{(s)}$ are invariant under Weyl rescaling of the $AdS_3$ radius $\ell$.
Supersymmetry imposes additional relations between the expansion coefficients
and, in particular, we find that
\begin{align}\label{d4sces}
X_{(s)}&=-\frac{L}{\ell}Y_{(1)}\,,\nn
Y_{(s)}&=Y_{(1)}\,.
\end{align}

We are interested in obtaining the expectation values for the scalar operators in flat space. We can follow
the discussion as for the $D=5$ case, but things are now simpler since there is no conformal anomaly in the $D=4$ setting.
After some calculation we find that with flat metric $ds^2=-dt^2+dy_1^2+dy_2^2$, and the spatial modulation in the $y_2$ direction,
the sources are given by
\begin{align}\label{d4sces2}
\frac{\ell^2}{y_2^2}X_{(s)}\,,\qquad
\frac{\ell}{y_2}Y_{(s)}\,,
\end{align}
with $y_2>0$ (since we are considering the $r\to\infty$ end).
Furthermore, the expectation values for ABJM theory are given by
\begin{align}\label{d4vevs}
\langle\mathcal{O}_{X}\rangle
&=\frac{7}{4\pi GL}\frac{\ell}{y_2}X_{(1)}\,,\nn
\langle\mathcal{O}_{Y}\rangle&=-\frac{L}{y_2}\langle\mathcal{O}_{X}\rangle\,.
\end{align}
Thus, for BPS configurations we see that for the ABJM side of an RG interface we can characterise both scalar sources using $Y_{(s)}$ 
and both expectation values can be determined from $X_{(1)}$.

For the $G_2^\pm$ side of the interface, things are simpler. First, holographic renormalisation for the $G_2^\pm$ fixed points does not
contain any local finite counterterms.
Next, recall that the complex scalar is dual to two operators 
$\mathcal{O}^{\Delta=1+\sqrt{6}}$ and $\mathcal{O}^{\Delta=2+\sqrt{6}}$. Since these are both irrelevant operators we need to demand
that their associated source terms vanish as $r\to -\infty$ in order that we approach the $G_2^\pm$ $AdS_4$ solution.
Examining the BPS equations we
then find that there is a universal mode associated with shifts in the coordinate $r$. Of more interest is that there is 
a mode of the form, as $r\to-\infty$,
\begin{align}\label{g2modes}
z&=z_{(G_2^\pm)}+(b-i)\zeta e^{r(1+\sqrt{6})/\LLS}+\dots\,
\end{align} 
that is parametrised by the real constant $\zeta$ and 
with $b$ real given by
\begin{align}
b =\pm\frac{\sqrt{3}+12 \sqrt{6}-\sqrt{30 \left(7 \sqrt{3}-12\right)}+9}{2 \sqrt{3} \left(-19 \sqrt{3}+\sqrt{310 \sqrt{3}-525}+95\right)^{1/2}}\,.
\end{align}
This mode is associated with the irrelevant operator $\mathcal{O}^{\Delta=1+\sqrt{6}}$ 
acquiring an expectation value in the $G_2^\pm$ theory.
More precisely, for this side of the interface at $r\to-\infty$, which is $y_2<0$ in the flat space boundary, using
\eqref{g2modes} we can define
\begin{align}\label{g2vevs1}
\langle\mathcal{O}^{\Delta=1+\sqrt{6}}\rangle
&\propto\left(\frac{\ell}{-y_2}\right)^{1+\sqrt{6}}\zeta\,.
\end{align}
The operator $\mathcal{O}^{\Delta=2+\sqrt{6}}$ also acquires an expectation value proportional to $\zeta$.

\subsubsection{The solutions}

\begin{figure}[h!]
\centering
{\includegraphics[scale=0.21]{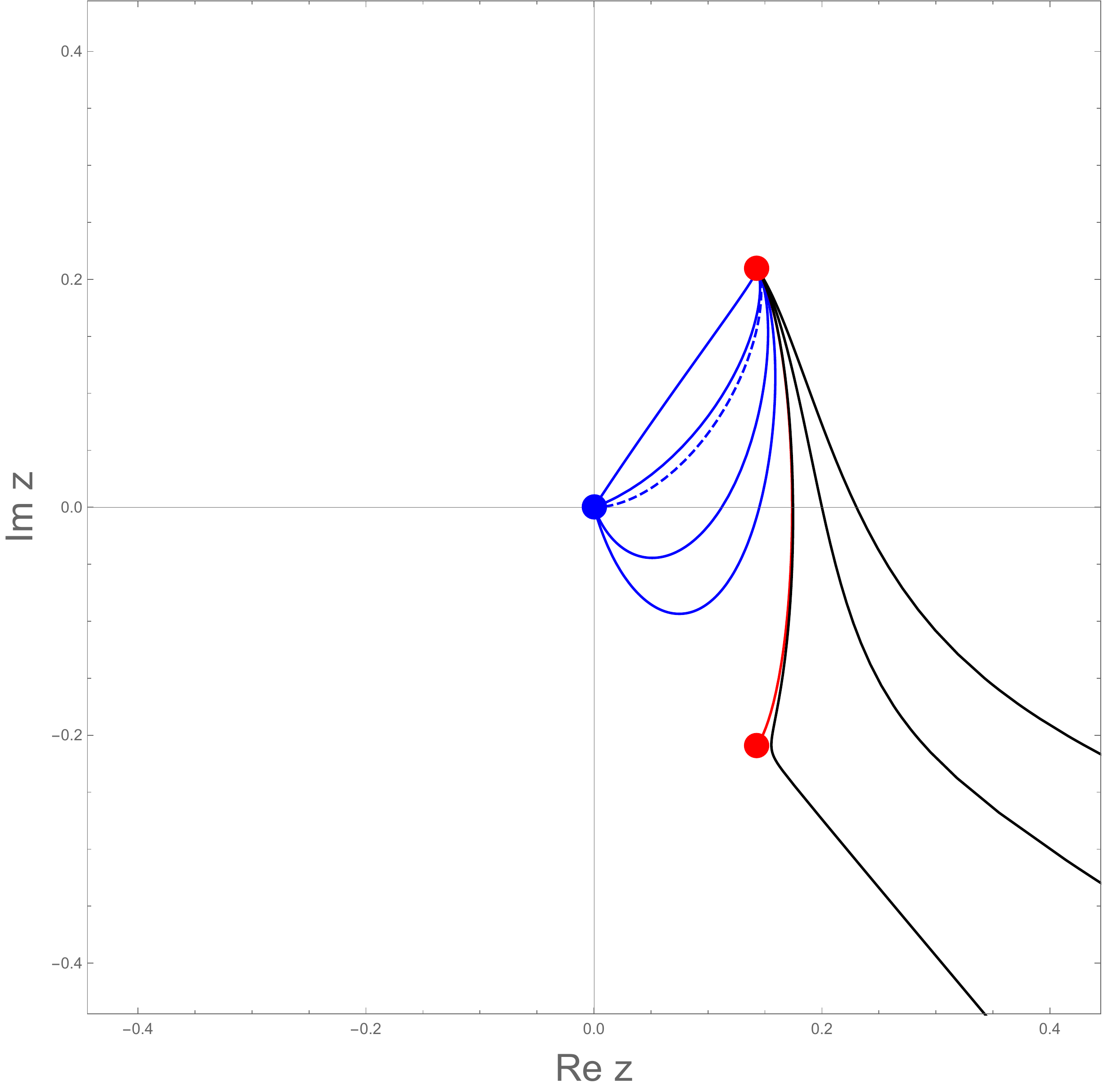}\qquad}\\
{\includegraphics[scale=0.23]{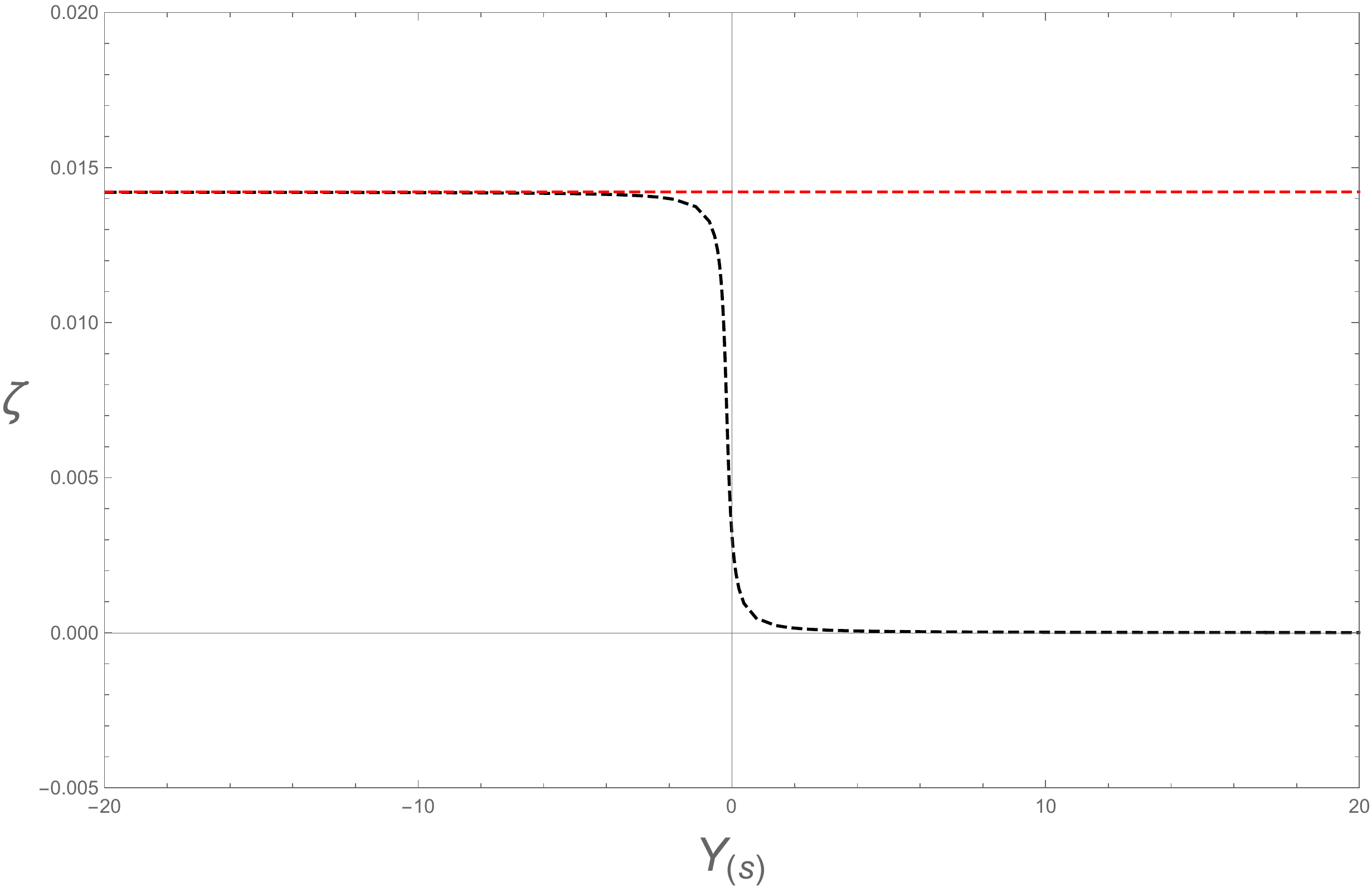}\qquad}
{\includegraphics[scale=0.21]{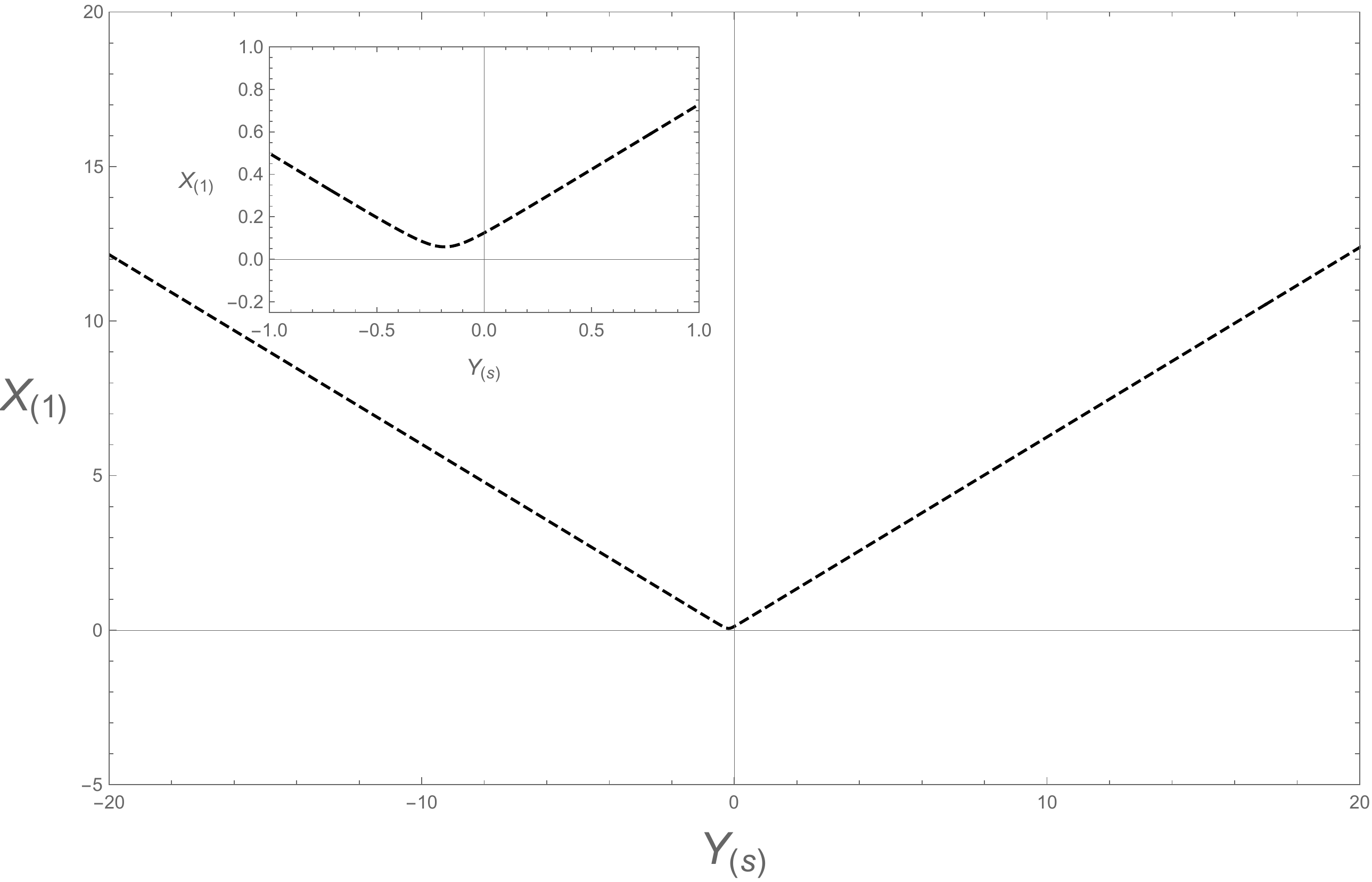}\qquad}
\caption{\label{fig:oned4} The first family of BPS solutions in $D=4$ is summarised by parametrically plotting the real and imaginary parts of 
the scalar field $z$ in the top panel. The blue dot is the ABJM $AdS_4$ solution and the two red dots
are the two $G_2^\pm$ $AdS_4$ solutions. 
The blue curves are dual to dual to ABJM/$G_2^+$ RG interfaces. For these solutions, the bottom panels plot
$\zeta$ and $X_{(1)}$, which fix the expectation values on the $G_2^+$ and ABJM side, respectively, as a function of $Y_{(s)}$ which fixes the
sources on the ABJM side.
The dashed blue line in the top panel is the RG interface solution
for which all sources vanish on the ABJM side of the interface. As $Y_{(s)}\to-\infty$ one approaches the $G_2^+$/$G_2^-$ Janus solution (red curve). The black curves become singular at $|z|=1$.
 }
\end{figure}

\begin{figure}[h!]
\centering
{\includegraphics[scale=0.2]{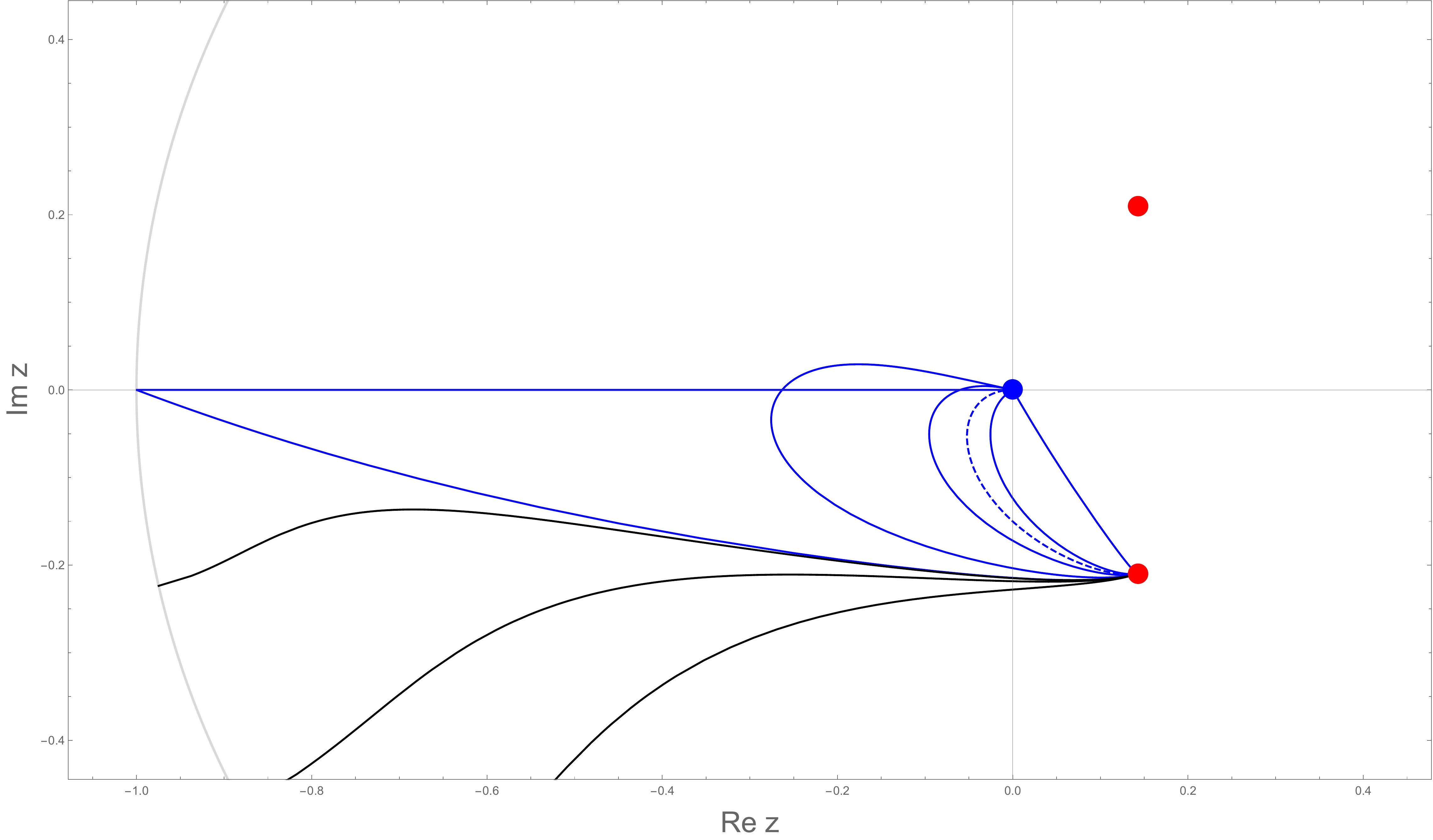}\qquad}\\
{\includegraphics[scale=0.24]{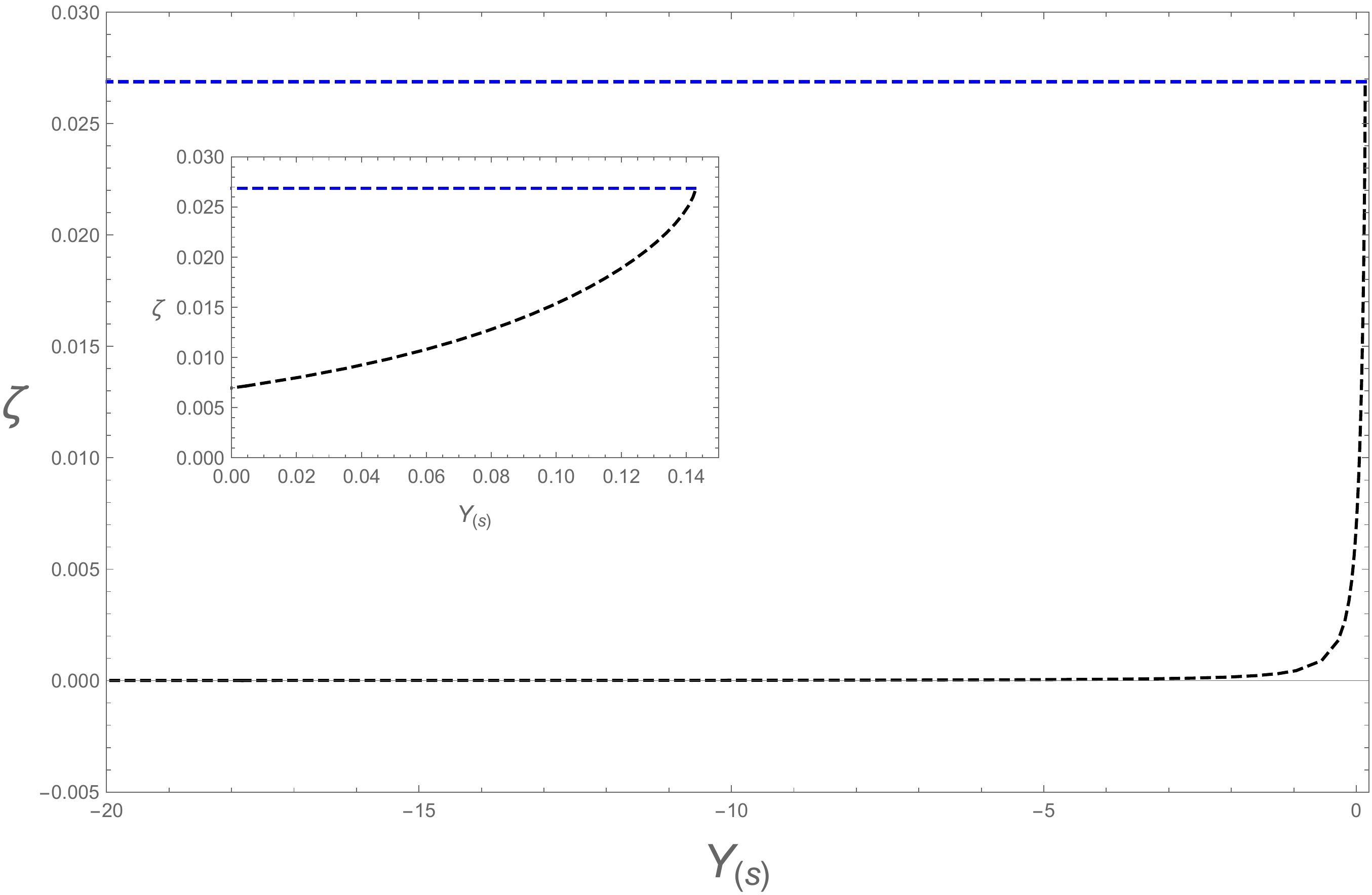}}\qquad
{\includegraphics[scale=0.23]{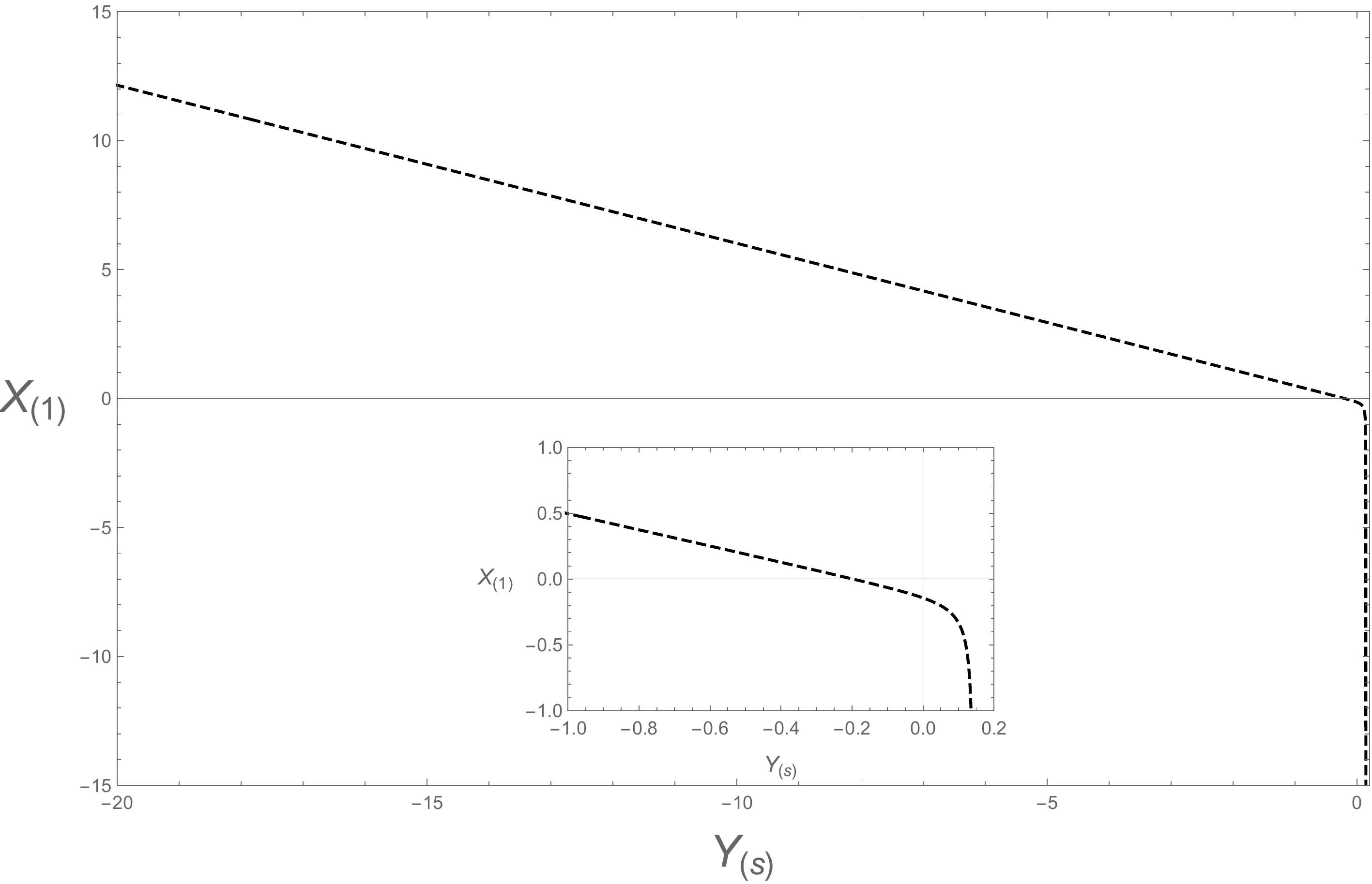}}
\caption{\label{fig:twod4} The second family of BPS solutions in $D=4$ presented analogously to the first family in figure \ref{fig:oned4}.
 The blue curves in the top panel are dual to ABJM/$G_2^-$ RG interfaces and the dashed blue line
 is the solution for which $Y_{(s)}=0$.
For the blue curves, the bottom panels plots the behaviour of $\zeta$ and $X_{(1)}$ which fix the expectation values on the $G_2^-$ side and ABJM side of the
RG interface, respectively, as a function of $Y_{(s)}$, which fixes the
sources on the ABJM side. In the limit $Y_{(s)}\to 0.1428$ one approaches the blue curve which hits $|z|=1$, the boundary of field space. In this family of solutions there is no
$G_2^+/G_2^-$ Janus solution.
 }
\end{figure}

We have numerically constructed two families of RG interface solutions by starting at either the $G_2^+$ or the $G_2^-$ fixed point at
$r=-\infty$, shooting out with the mode in \eqref{g2modes}, parametrised by $\zeta$ and giving rise to the 
expectation value \eqref{g2vevs1},
and then seeing where one ends up at $r=\infty$.
The main results are presented in figures \ref{fig:oned4} and \ref{fig:twod4}.
We note that we have set $\ell=1$ in \eqref{bpsd4}.

The family of solutions starting at the $G_2^+$ fixed point at $r\to-\infty$ are rather similar to those discussed in the previous section, so we discuss them first. Figure \ref{fig:oned4}, top panel, provides a parametric plot of the real and imaginary parts of the scalar field $z$ for this class of solutions. The blue dot at the origin is the $\mathcal{N}=8$ $AdS_4$ solution while the two red dots are the two $G_2^\pm$ $AdS_4$ solutions.
The blue curves are a one parameter family of RG interface solutions with ABJM theory on one side ($y_2>0$)
and $G_2^+$ on the other ($y_2<0$). The RG interface solutions exist in the range
$-\infty <Y_{(s)}< \infty$, where we recall $Y_{(s)}$ fixes all of the sources, and $0<\zeta<\zeta_{crit}$ with
$\zeta_{crit}\approx 0.01421$. When $Y_{(s)}\to -\infty$ (and $\zeta\to \zeta_{crit}$) the solutions approach the red curve and
when $Y_{(s)}\to +\infty$ (and $\zeta\to 0$) they approach a straight line connecting the $\mathcal{N}=8$ $AdS_4$ solution with the
$G_2^+$ $AdS_4$ solution.  
In the bottom panels of figure \ref{fig:oned4}, for the blue curves
we have plotted $\zeta$ and $X_{(1)}$, which give the expectation values on the $G_2^+$ and ABJM side of the interface, respectively, as a function of  $Y_{(s)}$. 
The general behaviour of the radial functions for all of the ABJM/$G_2^+$ RG interface solutions 
(blue curves in figure \ref{fig:oned4}) have a similar form; since they are somewhat analogous to figure \ref{fig:three} we don't
explicitly display them.

For this family of solutions we see in the bottom left panel of figure \ref{fig:oned4} that there is an RG interface solution for which 
 $Y_{(s)}=0$. This means that all sources on the ABJM side vanish, and since the sources always vanish on
 the $G_2^+$ side, remarkably this is an RG interface solution that has vanishing sources away from the interface.
For this special solution, marked by the 
dashed blue line in the top panel of figure 
\ref{fig:oned4}, we can determine the expectation values of the operators in the two SCFTS.
On the $G_2^+$ side we find $\zeta\approx 0.003184$. For the ABJM side, recalling
from \eqref{d4vevs} that the expectation values of the scalar operators are all determined by  
$X_{(1)}$, we find that $X_{(1)}\approx 0.1244$.

We next consider how the solutions behave as $Y_{(s)}\to +\infty$, with $\zeta\to 0$, when they approach a straight line between
the ABJM $AdS_4$ solution and the $G_2^+$ $AdS_4$ solution in figure \ref{fig:oned4}. Much as we saw in the $D=5$ case, in this limit
the solutions closely approximate two solutions joined together: there
is one region, on the ABJM side, that closely approach the Poincar\'e invariant RG flow
solution from the ABJM $AdS_4$ solution to the $G_2^+$ fixed point and this is joined together to another region which closely approximates
the $G_2^+$ fixed point solution itself, analogous to figure \ref{fig:four}.
From the $G_2^+$ side, as $Y_{(s)}\to +\infty$ we have $\zeta\to 0$ and this gives rise to the following critical exponent
\begin{align}\label{cexp1d4}
\zeta\sim  Y_{(s)}^{-\gamma},\qquad\qquad \gamma \approx1.449\,,
\end{align}
Recalling that $\zeta$ is giving the expectation value of an operator with conformal dimension $1+\sqrt{6}$ as in
\eqref{g2vevs1}, it seems highly likely that the exact critical exponent is $-1+\sqrt{6}$ and it would be interesting to prove this.

We now examine what happens to this family of RG interface solutions as $Y_{(s)}\to-\infty$, when we have
$\zeta\to \zeta_{crit}\approx  0.01421$, as in the bottom left
panel of figure \ref{fig:oned4}.
In this limit the blue curves in 
figure \ref{fig:oned4} approach the red curve which describes
a $G_2^+/G_2^-$ Janus solution with $G_2^+$ on one side of the interface and $G_2^-$ on the other.
Interestingly, on both sides of the interface of the Janus solution we have vanishing sources.
We can also show that this solution is invariant under the $\mathbb{Z}_2$ symmetry \eqref{zedtwo4}.

The way in which the RG interface solutions approach this $G_2^+$/$G_2^-$ Janus is also interesting.
As one might expect from figure \ref{fig:oned4}, the solutions closely approximate two solutions joined together: there
is one region, on the ABJM side of the interface ($r\to\infty$), that closely approaches
the Poincar\'e invariant RG flow solution from ABJM to the $G_2^-$ fixed point and this is joined
together to another region which closely approximates
the  $G_2^+$/$G_2^-$ Janus solution, analogous to figure \ref{fig:four}.
Focussing on the $G_2^+$ side of these limiting RG interface solutions we obtain
another critical exponent:
\begin{align}
\zeta_{crit}-\zeta\sim  |Y_{(s)}|^{-\gamma},\qquad\qquad \gamma \approx 1.450\,,
\end{align}
and again we believe this to be exactly $-1+\sqrt{6}$.

Figure \ref{fig:oned4} also shows that there is a one parameter family of solutions, the black curves, which approach $G_2^+$ as $r\to-\infty$, and then
approach a singular behaviour, with $|z|\to 1$, at some finite value of $r$. These solutions can be 
characterised by the expectation value of the operator in the $G_2^+$ SCFT and have $\zeta>\zeta_{crit}$, seemingly existing for arbitrary large values of $\zeta$. Although not plotted in figure \ref{fig:oned4} there are also singular solutions starting at  $G_2^+$ SCFT with $\zeta<0$ and hitting
a singularity at $|z|=1$,
which were also seen in \cite{Bobev:2013yra}. Here we have shown that these all describe configurations of the $G_2^+$ SCFT when placed on a half space without sources and with non-vanishing expectation values fixed by $\zeta$.

We now consider the second family of $D=4$ solutions, which are constructed by starting at the $G_2^-$ fixed point at $r\to-\infty$,
as summarised in figure \ref{fig:twod4}.
In the top panel, the blue curves are a one parameter family of RG interface solutions with ABJM theory on one side ($y_2>0$)
and $G_2^-$ on the other ($y_2<0$). In the bottom panel of figure \ref{fig:twod4} 
we have plotted $\zeta$ and $X_{(s)}$ which gives the expectation values on the $G_2^-$ and ABJM side of the interface, respectively,
as a function of  $Y_{(s)}$.
The RG interface solutions exist in the range
$-\infty <Y_{(s)}< 0.1428$ with $0<\zeta<\zeta_{crit}\approx 0.02688$.
When $Y_{(s)}\to 0.1428$ (and $\zeta
\to \zeta_{crit}$) the solutions approach the blue curve and horizontal line that hits the singularity at $|z|=1$.
When $Y_{(s)}\to -\infty$ (and $\zeta
\to 0$) they approach a straight line connecting the ABJM $AdS_4$ solution with the
$G_2^-$ $AdS_4$ solution. The general behaviour of the radial functions for all of the ABJM/$G_2^-$ RG interface solutions 
(blue curves in figure \ref{fig:one}) all have a similar form, somewhat analogous to figure \ref{fig:three}.

For this family of solutions, from the bottom panels of figure \ref{fig:twod4} we again find a source free RG interface solution for which 
 $Y_{(s)}=0$, marked by the dashed blue curve in the top panel of figure \ref{fig:twod4}.
For this solution, we find the following results for the expectation values of the operators in the two SCFTS.
On the $G_2^-$ side we find $\zeta\approx  0.006953$, while for the ABJM side we find that $X_{(1)}= -0.14312$.

We next consider how the solutions behave as $Y_{(s)}\to -\infty$ and $\zeta
\to 0$, when they approach a straight line between
the ABJM $AdS_4$ solution and the $G_2^-$ $AdS_4$ solution in figure \ref{fig:twod4}. Once again, in this limit
the solutions closely approximate two solutions joined together: there
is one region, on the ABJM side, that closely approach the Poincar\'e invariant RG flow
solution from the ABJM $AdS_4$ solution to the $G_2^-$ fixed point and this is joined together to another region which closely approximates
the $G_2^-$ fixed point solution itself, analogous to figure \ref{fig:four}.
Furthermore, from the $G_2^-$ side, as $Y_{(s)}\to -\infty$ we have $\zeta\to 0$ and this gives rise to the following critical exponent 
\begin{align}\label{cexp1d42}
\zeta\sim  |Y_{(s)}|^{-\gamma},\qquad\qquad \gamma \approx 1.450\,,
\end{align}
which we believe to be exactly $-1+\sqrt{6}$.

We now consider what happens to this second family of RG interface solutions as 
$Y_{(s)}\to 0.1428$, when we have
$\zeta\to \zeta_{crit}\approx 0.02688$, as in the bottom
left panel of figure \ref{fig:twod4}. In this limit the blue curves approach a singular solution that touches the boundary of field space at $|z|=1$.
As we approach the limiting solution, the solutions consist of two parts. One part is the curved line from the $G_2^-$ fixed point out to 
$|z|\to 1$, associated with the $G_2^-$ SCFT when placed on $AdS_3$ with radius $\ell=1$, without sources and with non-vanishing expectation values fixed by $\zeta$.
The second part is the horizontal blue line that goes to $|z|\to 1$. From the lower right plot in figure \ref{fig:twod4} we have $X_{(1)}\to-\infty$ as
$Y_{(s)}\to 0.1428$. We think that this part is approaching a Coulomb branch solution of ABJM theory on flat space with, after a suitable rescaling of $\ell$,
$X_{(1)}$ finite and $Y_{(s)}\to 0$.

Figure \ref{fig:twod4} also shows that there is again a one parameter family of solutions which approach $G_2^-$ as $r\to-\infty$, and then
approach a singular behaviour, with $|z|\to 1$, at some finite value of $r$. These solutions can be 
characterised by the expectation value of the operator in the $G_2^-$ SCFT and have $\zeta>\zeta_{crit}$, again seemingly existing for arbitrary large values of $\zeta$.  Although not plotted in figure \ref{fig:twod4} there are also singular solutions starting at  $G_2^-$ SCFT with $\zeta<0$ and hitting
a singularity at $|z|=1$ at finite $r$,
which were also seen in \cite{Bobev:2013yra}. Here we have shown that these all describe configurations of the $G_2^-$ SCFT when placed on a half space without sources and with non-vanishing expectation values fixed by $\zeta$.

\section{Discussion}\label{sec:disc}

In this paper we have constructed gravitational solutions that are dual to RG interface solutions and examined some of their properties.
In $D=5$ supergravity we found solutions dual to RG interface solutions with $\mathcal{N}=4$ SYM on one side and the $\mathcal{N}=1$ 
LS SCFT on the other. Generically, the solutions are supported by spatially dependent mass terms on the 
$\mathcal{N}=4$ SYM side of the interface, but there is one solution for which these vanish. 
As the source terms of the $\mathcal{N}=4$ SYM side diverge, we obtain a novel $D=5$ solution describing a LS$^+$/LS$^-$ Janus
solution. From the dual field theory point of view the Janus interface has the same LS SCFT on either side of the interface, but they are related
by the action of a discrete $R$-parity transformation, which is a novel feature.
The LS theory has a $CP$ symmetry and while the RG interface is not left invariant by the action of $R$ or $CP$ individually, it is
left invariant under their combined action.

In $D=4$ supergravity we extended and further studied the RG interface solutions found in \cite{Bobev:2013yra}.
These RG interfaces have ABJM theory on one side of the interface and one of the two $G_2$ invariant $\mathcal{N}=1$ $d=3$ SCFTs
on the other, which we argued are related by the action of a discrete $CP$ transformation.
We showed that the solutions are generically supported by spatially dependent mass terms on the 
ABJM side of the interface, but there is again one solution for which these vanish. 
As the source terms of the ABJM side diverge, we obtain a novel $D=4$ solution describing a $G_2^+/G_2^-$ Janus
solution. From the dual field theory point of view this RG interface has the two different 
$G_2$ invariant SCFTs on either side of the interface.

From the results of this paper
it seems likely that if a holographic Poincar\'e invariant RG flow solution from CFT$_{UV}$
to CFT$_{IR}$ exists then, generically, there will always be a corresponding RG interface solution. Indeed it is difficult to see why, generically, 
the existence of such a gravitational solution would be obstructed.
Furthermore, it seems likely that these RG interface solutions will be supported by spatially dependent sources on
the CFT$_{UV}$ side of the RG interface and vanishing sources on the CFT$_{IR}$ side, but there could be classes where
there are additional sources activated on the latter. 
It is natural to conjecture that there will always be a special solution for which the sources
away from the interface all vanish, as we have seen in the examples of this paper. 
It is also natural to expect that there will also be limiting Janus solutions when 
the CFT$_{IR}$ has a discrete automorphism, as in the LS case, 
or is related to another CFT by a discrete parity transformation, as in the $G_2^\pm$ case.

Additionally, it would be interesting to investigate setups for which there are Poincar\'e invariant RG flows from CFT$_{UV}$ to two IR CFTs, CFT$_{IR}$ and CFT$^\prime$$_{IR}$, which are not related by any parity trasnformation. For example, it may be possible to have situations for which there is no Poincar\'e invariant RG flow between CFT$_{IR}$ and CFT$^\prime$$_{IR}$, yet a conformal interface between the two still exists. 
In situations for which there is a Poincar\'e invariant RG flow between CFT$_{IR}$ and CFT$^\prime$$_{IR}$ then one can envisage
RG interfaces with multiple interfaces.

It would be interesting to explore these ideas further by explicitly constructing additional examples of type IIB and $D=11$ supergravity.
For example, we think it would be worthwhile to construct RG interface solutions separating the ABJM SCFT with the
$\mathcal{N}=2$ $d=3$ SCFT with $SU(3)\times U(1)$ global symmetry, for which the associated Poincar\'e invariant RG flows
have been constructed \cite{Ahn:2000aq,Bobev:2009ms}. It should be possible to construct various interface solutions, similar to those in this paper, 
using the consistent truncation discussed in \cite{Bobev:2010ib}.

In this paper we have elucidated what is happening to the sources and expectation values of various operators on either side of the interface, both for
the RG interface solutions and the Janus solutions. It would be interesting to further understand what is happening on the interface itself.
While this is somewhat delicate, we note that the distributional sources for a class of holographic supersymmetric Janus solutions were
explicitly determined in \cite{Arav:2018njv}. Although, the derivation of \cite{Arav:2018njv} utilised the fact that the BPS equations boiled down to solving the Helmholtz equation on the complex plane, we expect it should be possible to suitably generalise the analysis to the present setting.
It would be interesting to explore transport across the interface, analogous to what was recently done in the context of $d=2$ CFTs
using holography in \cite{Bachas:2020yxv}.

Finally, we have also discussed $D=5$ and $D=4$ solutions which are non-singular on one side of the interface, approaching the 
LS$^\pm$ $AdS_5$ or the $G_2^\pm$ $AdS_4$ solutions, respectively, and singular on the other.
Similar solutions were argued to be dual to BCFTs in \cite{Gutperle:2012hy}. We have shown that the singular solutions
have vanishing source terms on the non-singular side of the interface. It would be worthwhile to further investigate the nature of the
singularity in $D=10$ and $D=11$ supergravity in order to determine the precise dual interpretation of these solutions.

\subsection*{Acknowledgments}
We thank Nikolay Bobev for discussions.
This work was supported by STFC grant ST/P000762/1 and
with support from the European Research Council under the European Union's Seventh Framework Programme (FP7/2007-2013), ERC Grant agreement ADG 339140. 
KCMC is supported by an Imperial College President's PhD Scholarship. 
JPG is supported as a KIAS Scholar and as a Visiting Fellow at the Perimeter Institute. 
The work of CR is funded by a Beatriu de Pin\'os
Fellowship.

\appendix

\section{Holographic renormalisation for the $D=4$ gravity model}\label{holrena}
We consider the class of $D=4$ bulk metrics of the form
\begin{align}
ds_4^2=\gamma_{ab}dx^adx^b+dr^2\,,
\end{align}
where $\gamma_{ab}=\gamma_{ab}(r,x)$ and the conformal boundary is located at $r\to \infty$.
The renormalised action can be written as the sum of four terms
\begin{align}
S=S_{\text{bulk}}+S_{\text{GH}}+S_{\text{ct}}+S_{u=0}\,.
\end{align}
The first two terms are the bulk action and the standard Gibbons-Hawking term
\begin{align}
&S_{\text{bulk}}+S_{\text{GH}}=\frac{1}{8\pi G}\int drd^3x \sqrt{|g|}\mathcal{L}
+\frac{1}{8\pi G}\int d^3x\, \sqrt{|\gamma|}K\,,
\end{align}
where the bulk Lagrangian $\mathcal{L}$ is given in \eqref{bulklag}. The term $S_{\text{ct}}$ is a boundary counterterm action that contains
terms to cancel divergences as well as finite counterterms that are required for a supersymmetric renormalisation scheme.
By using the Bogomol'nyi trick of \cite{Freedman:2013ryh} we deduce that we should have
\begin{align}
S_{\text{ct}}&=\frac{1}{16\pi G}\int d^3x\, \sqrt{|\gamma|}\Big\{-4e^{\mathcal{K}/2}|\mathcal{V}|-LR\Big\}\,,
\end{align}
where here $R$ is the Ricci scalar of the metric $\gamma_{ab}$ as $r\to\infty$. The final boundary term $S_{u=0}$ is also needed for supersymmetry. Indeed writing the complex scalar as $z=X+iY$ it ensures that $X$ is dual to an operator with $\Delta=1$ (alternative quantisation) and $Y$ is dual to an operator with $\Delta=2$ (standard quantisation). Following the procedure of \cite{Cabo-Bizet:2017xdr} we need to carry out a suitable Legendre transformation
and we find
\begin{align}
S_{u=0}=\frac{1}{16\pi G}\int d^3x \sqrt{|\gamma|}\Big\{28X\partial_rX+\frac{28}{L}X^2+\frac{84}{L}X^3-\frac{84}{L}XY^2\Big\}\,,
\end{align}
evaluated at $r\to\infty$.

For a general class of solutions preserving $ISO(1,1)$ symmetry, with $\gamma_{ab}dx^adx^b=e^{2A(r,x)}(-dt^2+dy^2)+e^{2V(r,x)} dx^2$ and the scalar fields functions of $(r,x)$, we have calculated the stress tensor and shown that the Ward identities are explicitly satisfied. Furthermore, we have also shown that for this class of solutions, the energy density is a total spatial derivative as in the $D=4$ models in
\cite{Gauntlett:2018vhk,Arav:2018njv} and the $D=5$ models discussed in \cite{Arav:2020obl}.


\providecommand{\href}[2]{#2}\begingroup\raggedright\endgroup

\end{document}